\documentclass[a4paper,12pt]{article}
\usepackage[]{epsfig}
\usepackage[]{rotating}
\usepackage[]{cite}

\newcommand{\Date}      {\today}

\newcommand{\ee}{\mbox{${\mathrm{e}}^+ {\mathrm{e}}^-$}}
\newcommand{\tautau}{\mbox{$\tau^+\tau^-$}}

\newcommand{\gevcs}      {\mbox{$\mathrm{GeV}/c^2$}}
\newcommand{\tevcs}      {\mbox{$\mathrm{TeV}/c^2$}}
\newcommand{\mhmax}      {\mbox{$m_{\mathrm{h^0}}{\mathrm{-max}}$}}
\newcommand{\cls}        {\mbox{$\mathrm{CL_s}$}}
\newcommand{\clsb}       {\mbox{$\mathrm{CL_{s+b}}$}}
\newcommand{\clb}        {\mbox{$\mathrm{CL_b}$}}
\newcommand{\hwa}        {\mbox{$\mathrm{H^\pm}\ra\mathrm{W^{*\pm}A^0}$}}
\newcommand{\hpwa}       {\mbox{$\mathrm{H^+}\ra\mathrm{W^{*+}A^0}$}}
\newcommand{\qq}         {\mbox{$\mathrm{q}\bar{\mathrm{q}}$}}
\newcommand{\csbar}      {\mbox{$\mathrm{c}\bar{\mathrm{s}}$}}
\newcommand{\bb}         {\mbox{$\mathrm{b}\bar{\mathrm{b}}$}}
\newcommand{\cc}         {\mbox{$\mathrm{c}\bar{\mathrm{c}}$}}

\newcommand{\mZ}         {\mbox{$m_{\mathrm{Z^0}}$}}
\newcommand{\mH}         {\mbox{$m_{\mathrm{H^0}}$}}
\newcommand{\mh}         {\mbox{$m_{\mathrm{h^0}}$}}
\newcommand{\mA}         {\mbox{$m_{\mathrm{A^0}}$}}

\newcommand {\Ho}        {\mbox{$\mathrm{H}^{0}$}}
\newcommand {\Ao}        {\mbox{$\mathrm{A}^{0}$}}
\newcommand {\ho}        {\mbox{$\mathrm{h}^{0}$}}
\newcommand {\Zo}        {\mbox{$\mathrm{Z}^{0}$}}
\newcommand{\Hpm}{\mbox{$\mathrm{H}^{\pm}$}}
\newcommand{\Hp}{\mbox{$\mathrm{H}^+$}}
\newcommand{\Hm}{\mbox{$\mathrm{H}^-$}}

\newcommand{\tpnu}{\mbox{${\tau^+\nu_{\tau}}$}}

\newcommand{\mHpm}{\mbox{$m_{\mathrm{H}^{\pm}}$}}

\newcommand{\WW}         {\mbox{$\mathrm{W}^+\mathrm{W}^-$}}
\newcommand{\pb}         {\mbox{$\mathrm{pb}^{-1}$}}
\newcommand{\tanb}       {\mbox{$\tan\beta$}}
\newcommand{\msusy}       {\mbox{$M_{\mathrm{SUSY}}$}}
\newcommand{\mg}       {\mbox{$m_{\mathrm{\tilde g}}$}}

\def\etal{\mbox{{\it et al.}}}
\newcommand{\OPALColl}  {OPAL Collab.}
\newcommand{\ra}        {\mbox{$\rightarrow$}}   
\begin{document}
\begin{titlepage}
\centerline{\Large EUROPEAN ORGANIZATION FOR NUCLEAR RESEARCH}
\begin{flushright}
LHWG Note 2001-04 \\
ALEPH 2001-057 CONF 2001-037 \\
DELPHI 2001-114 CONF 537 \\
L3 Note 2700 \\
OPAL Technical Note TN699 \\      
\Date
\end{flushright}

\bigskip
\begin{center}{\LARGE\bf Searches for the Neutral Higgs Bosons of the MSSM:
Preliminary Combined Results Using LEP Data Collected at Energies up to 209~GeV}
\end{center}
\bigskip

\vspace{0.1cm}  
\begin{center}
 {\Large {The ALEPH, DELPHI, L3 and OPAL Collaborations, \\ and the LEP Higgs Working Group}} \\
\end{center}
\bigskip\bigskip

\bigskip
\begin{center}{\large  Abstract}\end{center}
In the year 2000 the four LEP experiments collected data at
centre-of-mass energies between 200 and 209~GeV, 
integrating approximately 870~\pb\ of luminosity, with about 510~\pb\ above 206~GeV.
The LEP working group for
Higgs boson searches has combined these data with data sets collected previously at lower
energies.  In representative scans of the parameters of the Minimal Supersymmetric Standard
Model (MSSM), the mass limits $\mh>91.0$~\gevcs\ and $\mA>91.9$~\gevcs\ are obtained for the light
CP-even and the CP-odd neutral Higgs boson, respectively.  For a top quark mass less than
or equal to 174.3~\gevcs, assuming that the stop quark mixing is maximal, and choosing 
conservative values for other SUSY parameters affecting the Higgs sector, the range
$0.5<\tanb<2.4$ is excluded.  Additionally, the results of flavour-independent searches for hadronically
decaying Higgs bosons are included, allowing exclusion of MSSM models with suppressed decays
of the Higgs bosons to pairs of b~quarks.

\bigskip\bigskip\bigskip\bigskip
\begin{center}
  {\Large\bf All results quoted in this note are preliminary.}

\end{center}

\end{titlepage}

%
%
%
%
%
%
\newpage
\section{Introduction}

This note describes a combination of preliminary results of searches
for the neutral Higgs bosons of the Minimal Supersymmetric Standard
Model (MSSM) by the four LEP collaborations, ALEPH, DELPHI, L3 and OPAL.
The results are based on data collected at $\sqrt{s} \approx 200 - 209$ GeV, 
the highest \ee\ collision energies attained at LEP, and are combined
with data collected earlier at lower centre-of-mass energies.

The MSSM predicts the existence of two complex scalar field doublets, with a total
of eight degrees of freedom.  As in the Standard Model (SM), three
degrees of freedom appear as the longitudinal polarization states of the
gauge bosons W$^+$, W$^-$ and \Zo.  The remaining five degrees of freedom
are manifested in five physical scalar Higgs states.  In this note,
the Higgs sector of the MSSM is assumed to conserve CP.  Under this assumption,
the physical Higgs bosons are the CP-even \ho\ and \Ho, the CP-odd \Ao, and the
charged bosons H$^+$ and H$^-$.  The quartic self-coupling of the Higgs fields
are determined by the gauge couplings, which limits the mass of the lighter of
the two CP-even Higgs bosons to be less than the mass of the \Zo\ at tree level.
Radiative corrections, particularly from loops containing the top quark, allow
the lightest Higgs boson mass to range up to approximately 135~\gevcs, which is its
maximal value~\cite{MSSMMHBOUND1,MSSMMHBOUND2,MSSMMHBOUND3,MSSMMHBOUND4,MSSMMHBOUND5,MSSMMHBOUND6,MSSMMHBOUND7,MSSMMHBOUND8}
 for all choices of parameters in the
MSSM models within the constrained framework considered in this note
(see Section~\ref{sect:benchmark}).
This constraint on the mass of the \ho\ suggests that the \ho\ may be
light enough to be produced at LEP.  Searches are performed for the possible
final states containing Higgs bosons and they are combined among the four
collaborations in order to place the tightest constraints on the possible values
of the parameters of the MSSM Higgs sector.

In the MSSM, the Higgs-strahlung process \ee\ra\ho\Zo\ proceeds as in the
Standard Model, but its rate is suppressed by the factor $\sin^2(\beta-\alpha)$, 
where $\tan\beta$ is the ratio of the vacuum expectation values of the two
field doublets, and $\alpha$ is the mixing angle in the neutral CP-even Higgs boson sector.
The WW- and ZZ-fusion processes of the SM also
proceed with a rate suppressed by the same factor relative to the SM rate.
Heavy-Higgs-strahlung, \ee\ra\Ho\Zo, occurs if it is kinematically possible,
and has the SM production cross-section suppressed by the factor $\cos^2(\beta-\alpha)$.
In some cases \ee\ra\Ho\Zo\ can have a higher cross-section than \ee\ra\ho\Zo.  The 
process \ee\ra\ho\Ao\ also occurs when kinematically allowed, and its production
cross-section is proportional to $\cos^2(\beta-\alpha)$.  
Dedicated analyses are used to search for this final state.

The Higgs boson sector of the MSSM corresponds to
a Type II two-Higgs-doublet model, in that the couplings of the Higgs
fields to the fermions are constrained such that at tree level one Higgs field couples to the
up-type fermions and the other to the down-type fermions and the charged leptons.  This
is arranged in order to avoid loop anomalies, to prevent flavour-changing neutral
currents, and to give mass to the up-type and down-type fermions.
This structure also implies that
the decay branching ratios of the Higgs bosons to fermions depend not only on
the masses, but also on the values of $\alpha$ and $\beta$.  At tree level, the coupling of the
\ho\ to \bb\ is proportional to $-\sin\alpha/\cos\beta$, the coupling
of the \ho\ to \cc\ is proportional to $\cos\alpha/\sin\beta$, the coupling
of the \Ao\ to \bb\ is proportional to $\tan\beta$, and the coupling of
the \Ao\ to \cc\ is proportional to $\cot\beta$.  Over much of the
parameter space considered, the \ho\ and the \Ao\ decay predominantly into
\bb\ and \tautau\ pairs, although for various choices of parameters,
the decays \ho\ra\Ao\Ao, \ho\ra\cc, \ho\ra gg and \ho\ra\WW\ can become important.

The searches that are combined in this note are the searches for the \ee\ra\ho\Zo\ 
(and WW- and ZZ-fusion) processes
which are used in the Standard Model interpretations presented separately by the four
collaborations in~\cite{ALEPHSMLETTER,DELPHISMLETTER,L3SMLETTER,OPALSMLETTER}, combined
with the searches for the \ee\ra\ho\Ao\ process described 
in~\cite{ALEPHMSSMMORIOND,DELPHIMSSMEPS,L3MSSMMORIOND,OPALMSSMMORIOND}.
In addition, for models in which the decay branching ratios of the Higgs bosons
to \bb\ and \tautau\ are suppressed, the flavour-independent results are 
used~\cite{ALEPHFLAVINDEP,DELPHIFLAVINDEP,L3FLAVINDEP,OPALMSSMMORIOND,OPALFLAVINDEP,LEPFLAVINDEP}.
In all combinations, a full specification of the production cross-sections
at all relevant centre-of-mass
energies and all decay branching ratios are incorporated into the calculations of
the expected signal rates.  The searches combined are sensitive predominantly to the
\bb\ and \tautau\ decays of the \ho\ and the \Ao.  A number of the searches mentioned
above also have estimated efficiencies for the decays \ho\ra\cc, gg, \WW, \Ao\Ao\ (with
specified decays of the \Ao, usually only to \bb ), etc.
The signal estimations for these searches also include the contributions from these
sources.  The Higgs boson masses, cross-sections and decay branching ratios
are computed using HZHA03~\cite{HZHA}, modified to use either the FeynHiggs 
calculations~\cite{MSSMMHBOUND7,weigheiholl,feynhiggs}
or SUBHPOLE2~\cite{MSSMMHBOUND5}.

Each experiment has generated Monte Carlo simulations of the signal processes
and the SM background processes, typically at centre-of-mass energies of
200, 202, 204, 206, 208 and 210~GeV.
The rates and distributions for energies in between the Monte Carlo points are interpolated.

The statistical procedure adopted for the combination of the data and the definitions
of the test statistic $-2\ln Q$ and the 
confidence levels \cls, \clsb\ and \clb, are described in \cite{LEPHIGGSEPS2001SM}.  The
main sources of systematic uncertainty in the estimations of the accepted signal and
background rates are incorporated using an extension of the method of Cousins and 
Highland~\cite{ref:cousinshighland}, where the correlations arising from shared error sources
between analyses conducted at different energies, and between similar analyses conducted
by the separate collaborations, are taken into account.

Searches for charged Higgs bosons are presented separately in~\cite{LEPHIGGSCHARGED}.

\section{Searches for {\boldmath{\ee\ra\ho\Ao}}}

The analyses of the full data sample for the \ho\Zo\ processes are 
documented in~\cite{ALEPHSMLETTER,DELPHISMLETTER,L3SMLETTER,OPALSMLETTER}.
  This section describes only
the results of searches for \ee\ra\ho\Ao.  In the MSSM, $\cos^2(\beta-\alpha)$ is
significantly different from zero only when $\tan\beta$ is large ($>8$) and \mA\ is less
than the maximum allowed \mh\ (depending on the parameters of the scenario).  The
models with $\cos^2(\beta-\alpha)$ significantly different from zero have
\mh$\approx$\mA.  The \bb\ and \tautau\ decays
of the \ho\ and \Ao\ are dominant for such models, 
and the searches concentrate on these
decays only.  The numbers of selected events,
the expected signal for \mh=90~\gevcs\  and \mA=90~\gevcs, and the estimated background from
SM processes are shown in Table~\ref{tab:ahevents}, separately for each experiment.
Also listed are the integrated luminosities reported by the experiments
for the data taken in the year 2000.  Due to the $\beta^3$ kinematical factor dependence of
the production cross-section for \ee\ra\ho\Ao, 
the expected limits on the \mh=\mA\ diagonal are 10~\gevcs\  below the
average energy of the LEP2 data in 2000, and so the precise distribution of the beam energy
is of less importance to the sensitivity of the \ho\Ao\ searches 
than it is to the \ho\Zo\ searches.  

\begin{table}[htbp]
\begin{center}
\begin{tabular}{|l|c|c|c|c|}\cline{2-5}
\multicolumn{1}{l|}{} & {\bf ALEPH} & {\bf DELPHI} & {\bf L3} & {\bf OPAL} \\ \hline
\multicolumn{5}{|c|}{{\bf\boldmath\ho\Ao\ra\bb\bb\unboldmath\ channel\rule{0mm}{5mm}}} \\ \hline
Integrated Luminosity (\pb ) & 217 & 224 & 217 & 208  \\
Data                         & 10  & 5   & 13  & 11   \\
Total Background             & 5.5 & 6.5 & 9.4 & 10.3 \\
Four-Fermion Bkg.            & 4.2 & 4.4 & 7.3 & 6.9  \\
\qq\ Background              & 1.4 & 2.1 & 2.1 & 3.4  \\
Efficiency                   & & & & \\
\mh = \mA = 90~\gevcs\       & 47\% & 47\% & 42\% & 48\% \\
Expected signal              & & & & \\
\mh = \mA = 90~\gevcs\       & 3.5 & 3.6 & 3.2 & 3.4 \\ \hline\hline
\multicolumn{5}{|c|}{{\bf\boldmath\ho\Ao\ra\bb\tautau\unboldmath\ channel\rule{0mm}{5mm}}} \\ \hline
Integrated Luminosity (\pb ) & 217 & 224 & 217 & 205 \\
Data                         & 3   &   5 & 2   &   5 \\
Total Background             & 3.0 & 6.0 & 3.0 & 4.5 \\
Four-Fermion Bkg.            & 2.8 & 5.6 & 2.8 & 4.1 \\
\qq\ Background              & 0.2 & 0.4 & 0.4 & 0.4 \\
Efficiency                   & & & & \\
\mh = \mA = 90~\gevcs\       & 41\% & 25\% & 33\% & 43\% \\
Expected signal              & & & & \\
\mh = \mA = 90~\gevcs\       & 0.6 & 0.4 & 0.4 & 0.6 \\ \hline\hline
Limit obs (exp.med) for \mh\ (\gevcs ) & 89.6 (91.7) & 89.7 (88.8) & 83.2 (88.1) & 79.3 (85.1) \\
Limit obs (exp.med) for \mA\ (\gevcs ) & 90.0 (92.1) & 90.7 (89.7) & 83.9 (88.3) & 80.6 (86.9) \\\hline
\end{tabular}
\end{center}
\vspace*{-0.5cm}
\caption[]{\label{tab:ahevents}
The results in the \ho\Ao\ channels for each experiment for the data taken
in 2000.  Listed are the individual signal efficiencies, the expected signal
counts, the total backgrounds, the backgrounds broken down into \qq\ and four-fermion
sources  and the observed data counts, for each experiment's \ho\Ao\ra\bb\bb\ and 
\ho\Ao\ra\bb\tautau\ channel separately.  The ``tight selection'' is shown
for DELPHI's \ho\Ao\ra\bb\bb\ channel for easier comparison with the
other experiments.  The L3 results are also shown with tighter selections than
are used in the combinations for easier comparison.
The signal efficiencies and rates are given for
$\mh =\mA =90$~\gevcs, with $\tan\beta\sim 20$.  Also listed are the observed and
median expected lower bounds on \mh\ and \mA, taking the lower
values of the limits obtained in the 
no-mixing and \mhmax\ scenarios.  These scenarios are
discussed in Section~\protect\ref{sect:mssmlimit}.}
\end{table}

\section{Limits in the MSSM Parameter Space}\label{sect:mssmlimit} 

The \ho\Zo\ and \ho\Ao\ searches at LEP in the year 2000
are combined with previous LEP Higgs searches presented in~\cite{LEPHIGGS202} and
references therein, conducted at centre-of-mass energies between $\sim 88$~GeV  and 202~GeV.

\subsection{Benchmark Scenarios}\label{sect:benchmark}
We test for the presence of an MSSM Higgs boson signal using a constrained 
model with seven parameters, \msusy, $M_2$, $\mu$, $A$, $\tanb$, \mA\ and \mg.
All of the soft SUSY-breaking parameters 
in the sfermion sector are set to \msusy\ at the electroweak scale.
$M_2$ is the SU(2) gaugino mass parameter at the electroweak scale, and $M_1$
is derived from $M_2$ using the GUT relation 
$M_1=M_2(5\sin^2\!\theta_W/3\cos^2\!\theta_W)$,
where $\theta_W$ is the weak mixing angle\footnote{$M_3$, $M_2$ and $M_1$ are
the mass parameters associated with the SU(3), SU(2) and U(1) subgroups
of the Standard Model.  The relevance of $M_3$ only enters via 
loop corrections sensitive to the gluino mass.}.  
The supersymmetric
Higgs boson mass parameter is denoted $\mu$, and \tanb\ is the ratio of the vacuum expectation
values of the two Higgs field doublets.  
The parameter $A$ is the common trilinear
Higgs-squark coupling parameter, assumed to be the same for up-type squarks
and for down-type squarks.  The largest contributions to \mh\ from radiative
corrections arise from stop loops, with much smaller contributions from
sbottom loops.  The gluino mass \mg\ affects loop corrections from both stops and
sbottoms.  The mass of the top quark is assumed to be 174.3~\gevcs, the
current average~\cite{RPP2000} of the TeVatron measurements.

Three benchmark scenarios
are considered~\cite{newbenchmarks}.  The first (``no-mixing'' scenario)
assumes that there is no
mixing between the scalar partners of the left-handed and the right-handed
top quarks, with the following values and ranges
for the parameters: $\msusy = 1$~\tevcs, $M_2 = 200$~\gevcs,
$\mu=-200$~\gevcs, $X_t(\equiv  A - \mu\cot\beta) =0$, $0.4<\tanb < 50$ 
and 4~\gevcs~$<\mA < 1$~\tevcs.  The
gluino mass \mg\ is set to 800~\gevcs; it has little effect on the phenomenology
of this scenario.
Most of the experimental Monte Carlo samples
assume that the \ho\ and \Ao\ have decay widths which are small compared to the
resolutions of the reconstructed masses; only DELPHI has performed 
tests in which the \ho\ and \Ao\ widths are significant~\cite{DELPHIMSSMEPS}.
The assumption that the decay widths can be neglected is only valid for
$\tan\beta<30$ in this scenario, and hence higher values of $\tan\beta$ are
not considered.  

The second scenario (``\mhmax '') is designed to yield the
maximal value of \mh\ in the model.
The \mhmax\ scenario corresponds to the most conservative range of excluded \tanb\ values
for fixed values of the mass of the top quark and \msusy.  The dependence of the
limit on \tanb\ on the top quark mass is given in Section~\ref{sect:resmhmaxnomix}.
The values of the parameters in the \mhmax\ scenario are fixed at the same
values used in the no-mixing scenario, except for the stop mixing parameter
$X_t = 2\msusy$, using the conventions of the two-loop diagrammatic calculation
of~\cite{MSSMMHBOUND7,weigheiholl}.  Only values of $\tan\beta$ below 30 are considered
in this model also in order to satisfy the assumptions made on the decay widths.

The third scenario (``large $\mu$'' scenario) is a scan with parameters
chosen to be $\msusy=400$~\gevcs, $\mu=1$~\tevcs,
$M_2=400$~\gevcs, $\mg=200$~\gevcs, $4\le\mA\le 400$~\gevcs, $X_t = -300$~\gevcs.
This third scenario is designed to illustrate choices of MSSM parameters
for which the Higgs boson \ho\ does not decay into pairs of b quarks
due to large corrections from SUSY loop processes.  This situation occurs mostly at
$\tanb > 20$ and for $120 < \mA < 220$~\gevcs.  The dominant decay modes
of the \ho\ for these models are to \cc, gg, \WW\ and \tautau.  For many of
these models, the decay \ho\ra\tautau\ is also suppressed, providing an additional
experimental challenge.  In this scenario, for all choices of \mA\ and \tanb, 
at least one Higgs boson signal with a large production cross-section
is within the kinematic reach of LEP2.  The maximum value of \mh\ in this scenario
is slightly less than 108~\gevcs.  For some choices of \mA\ and \tanb, the
\ho\Zo\ cross-section is suppressed by a small value of $\sin^2(\beta-\alpha)$,
and \ee\ra\ho\Ao\ is not within the kinematic reach of LEP2.  For these models,
however, the heavy Higgs \Ho\ has a mass less than 109~\gevcs, and may be
produced in Higgs-strahlung.  The value of Br(\Ho\ra\bb) may be suppressed, however.
For all choices of parameters within the recommended ranges
in the large~$\mu$ scenario, 
the decay widths of the \ho\ and the \Ao\ remain small when compared with
the resolutions on the reconstructed masses.  For this reason,
the full recommended range of $\tan\beta$ up to 50 is considered in
this scenario, in contrast to the first two scenarios.

For the no-mixing and \mhmax\ scenarios, the two-loop diagrammatic
approach of~\cite{MSSMMHBOUND7,weigheiholl} is used to compute the relations between
the SUSY parameters, \mh, \mA, \mHpm, \tanb, and the production cross-sections
and decay branching ratios.  For the large $\mu$ scenario, the
one-loop renormalization-group improved calculation 
of~\cite{MSSMMHBOUND5,carenamrennawagner,reconciliation}
is used.  These two calculations give consistent 
results~\cite{reconciliation,espinosareconciliation},
although small differences still exist.  For example, in the \mhmax\  
scenario, the diagrammatic approach gives a more conservative
upper edge of the excluded region of
\tanb, while one-loop renormalization-group improved approach gives
a slightly more conservative lower edge.
The value of \mh\ predicted by FeynHiggs
is uncertain at the 2--3~\gevcs\ level, due to uncalculated subleading and
higher-order corrections.  A recent calculation is available~\cite{MSSMMHBOUND8} which
incorporates subleading two-loop top-Yukawa terms not yet included in FeynHiggs,
although its applicability is limited to situations with $\mA\gg\mZ$.  The differences
with the FeynHiggs calculation fall within the 2--3~\gevcs\ uncertainty, and are
smaller than those induced by the current uncertainty in the top quark mass. The
effect of the uncertain top quark mass on the \tanb\ limit is given below.

\subsection{Results}

The calculations of the confidence levels are performed for the three benchmark scenarios
separately, and the results are shown in this section.  The \mhmax\ and no-mixing
scenarios are described together in Section~\ref{sect:resmhmaxnomix}, while the
large~$\mu$ scenario is described separately in Section~\ref{sect:reslargemu} because
of qualitative differences in the features of these scenarios.

\subsubsection{The {\boldmath\mhmax\unboldmath} and No-Mixing Scenarios}
\label{sect:resmhmaxnomix}

\begin{figure}[p]
\centerline{\epsfig{file=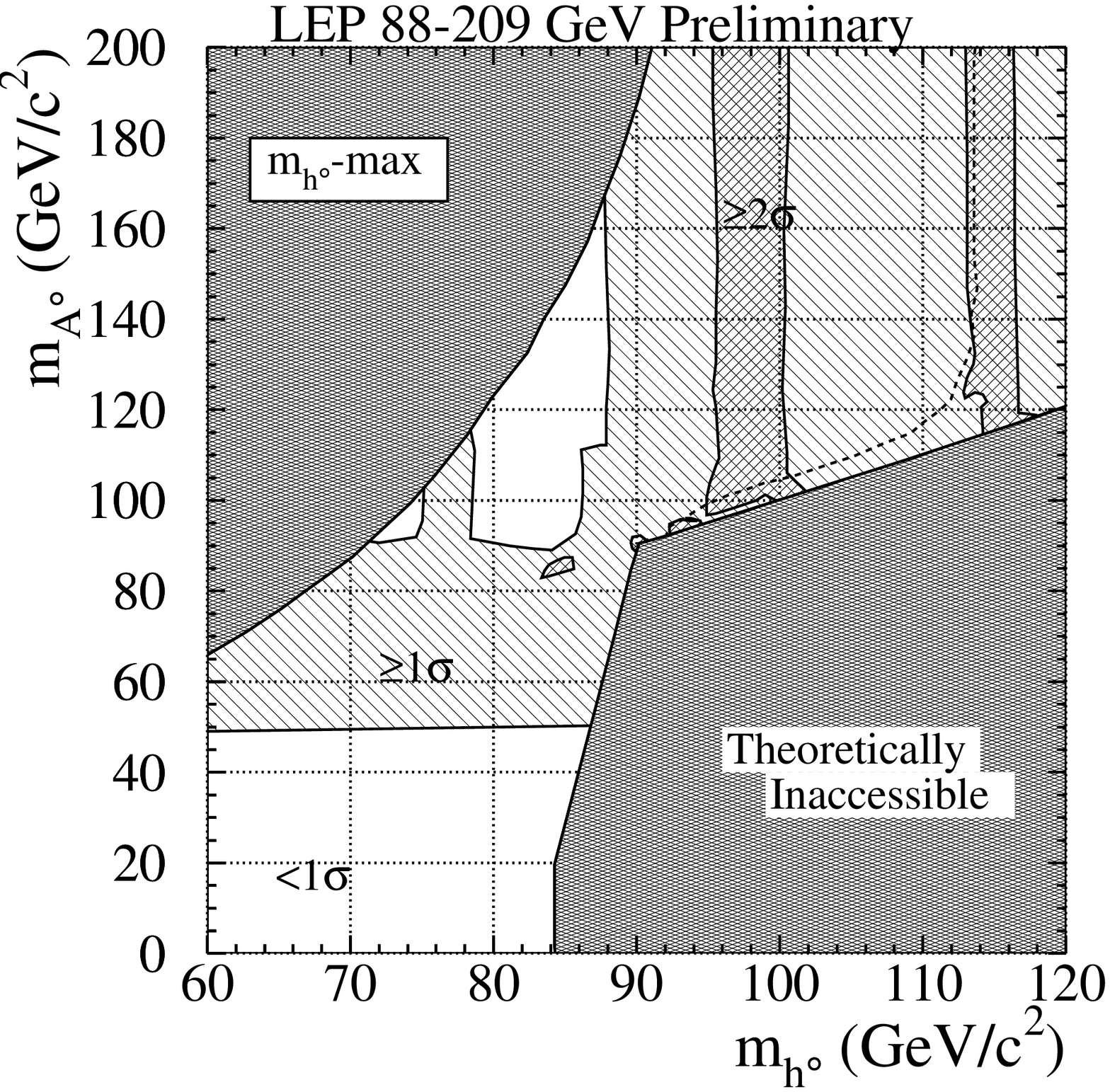,width=0.9\textwidth}}
\caption[]{\label{fig:mssm_clb}
The distribution of the confidence level ${\mathrm{CL}}_b$ in the (\mh,~\mA) plane
for the \mbox{\mh -max} scenario.  In the white domain, the observation either
shows a deficit or is less than 1$\sigma$ above the background 
prediction,
while in the domains labelled $\ge 1\sigma$ and $\ge 2\sigma$, the
observation shows an excess above the SM background prediction 
($1-\clb <0.32$, $1-\clb < 0.05$, respectively).  
If at a point (\mh,~\mA) in the plane, two values of $\tan\beta$ are
allowed by the benchmark model, the choice of $\tan\beta$ with
 the smaller $1-\clb$ is shown.
Results from the \ho\Zo\ searches are combined with the results of the
\ho\Ao\ searches.  Vertical structures are due to features in the \ho\Zo\ search
results, while structure on the \mh=\mA\ line arises from the \ho\Ao\ searches.
The 95\% CL exclusion contour is shown with the
dashed line; points to the right and below the dashed line
are unexcluded.  These regions can also be seen in Figure~\protect\ref{fig:maxmh}.
}
\end{figure}

Figure~\ref{fig:mssm_clb} shows the $1-\clb$ significance contours 
as functions of $\ho$ mass and $\Ao$ mass for the \mhmax\ scenario.
An excess is seen at $(\mh,\mA)\sim(83,83)$ \gevcs, with a significance level
slightly in excess of 2$\sigma$.  This is due to candidates in the OPAL 189~GeV
\tautau\bb\ channel~\cite{pr189} which have not been confirmed by later running or in other
experiments; the significance has gradually decreased as additional luminosity
has been accumulated.  Another excess is seen near $(\mh,\mA)\sim(93,93)$ \gevcs,
due to candidates in the OPAL four-jet channel in the data taken
in 2000~\cite{OPALMSSMMORIOND}, which also
does not appear in other samples.  The current 95\% CL exclusion limits from
LEP (shown also in the same figure) rules out the possibility of a signal
with $(\mh,\mA)\sim(83,83)$ \gevcs\ within the MSSM models considered, but is not
strong enough to exclude the $(\mh,\mA)\sim(93,93)$ \gevcs\ hypothesis.
There are two excesses in the \ho\Zo\ searches which appear as vertical bands
in Figure~\ref{fig:mssm_clb}, at $\mh\approx 97$~\gevcs, and at $\mh\approx 115$~\gevcs.
The excess at \mh=97~\gevcs\ is present in the 189~GeV data collected in 
1998~\cite{LHWGTAMPERE}, but does not appear in the 192--202~GeV data collected
in 1999~\cite{LEPHIGGS202}. Its significance is only slightly over 2$\sigma$.
The excess at $\mh\approx 115$~\gevcs\ is discussed in ~\cite{LEPHIGGSEPS2001SM};
its significance is also only slightly over 2$\sigma$.

Due to the large range of models investigated and the fine reconstructed
mass resolutions, the probability to have a 2$\sigma$ excess somewhere is much larger
than the 5\% it would be if only a single counting experiment had been done.
Over the range shown in Figure~\ref{fig:mssm_clb}, the dilution factor of the significance
is estimated to be 30--60.  This estimation was performed by scaling the signal and background
estimations of DELPHI and OPAL's test-mass-independent analyses of the 1999 and earlier 
data\footnote{Since this estimation was done, OPAL has created new test-mass-dependent
analyses for the 1999 data.} by a factor
of two, randomly generating candidates according to the background estimations, and for
each set of random candidates, performing a confidence level calculation in the
\mh-max\ scenario and noting the smallest $1-\clb$ obtained.  The probability of
obtaining a particular value of $1-\clb$ or smaller was estimated and compared against
$1-\clb$.  More than one independent 2$\sigma$ excess is probable.

\begin{figure}[p]
\centerline{\epsfig{file=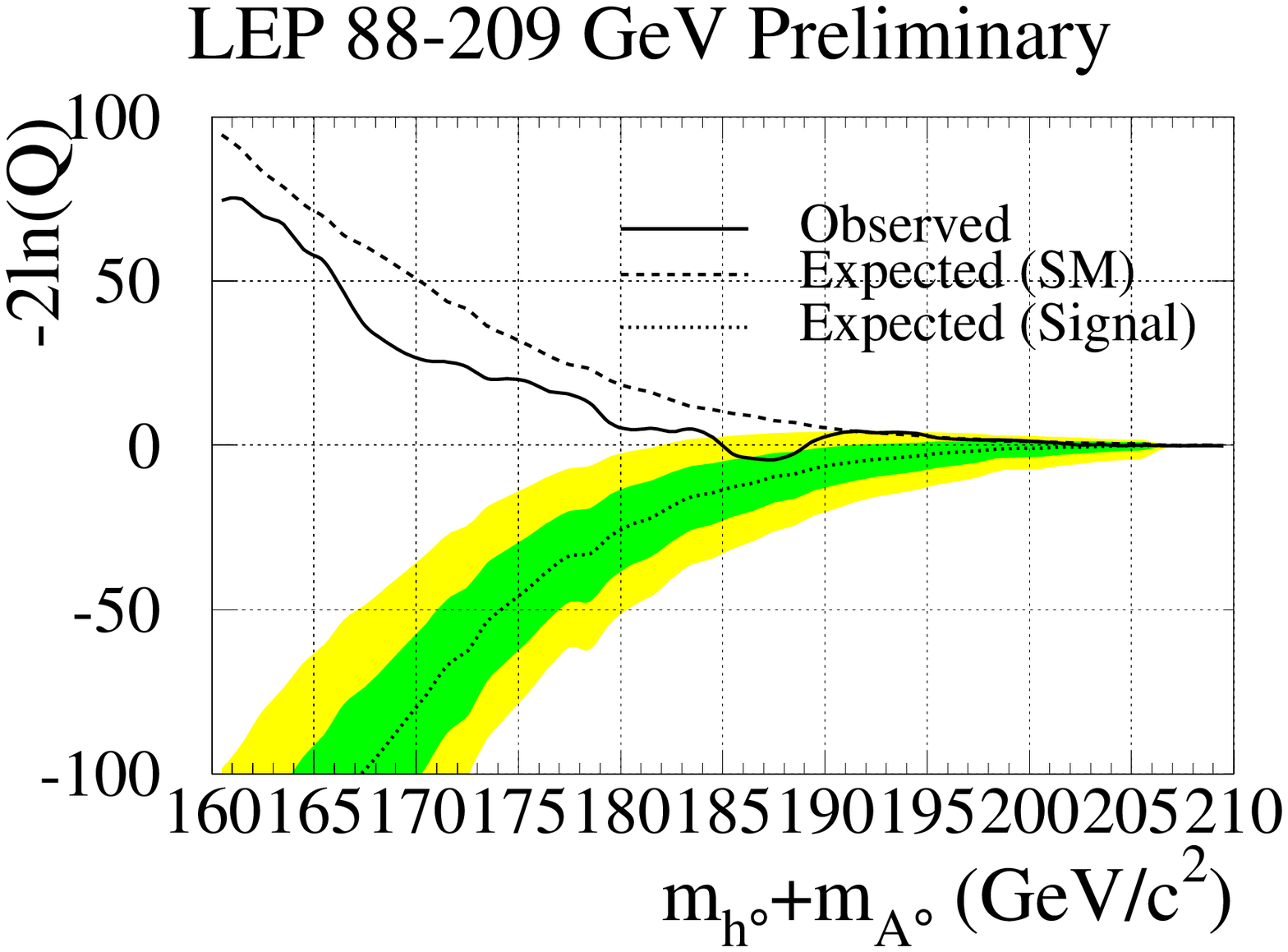,width=0.95\textwidth}}
\caption[]{\label{fig:diag2lnq}  
The value of the likelihood ratio $-2\ln Q$ as a function of the
sum of the test masses \mh + \mA, for $\mh\approx\mA$ 
($\tan\beta>20$, $\cos^2(\beta-\alpha)\approx 1$) in the \mhmax\ scenario.
The solid curve shows the observed values of -2ln$Q$ in the combination of the
four experiments' results; the upper dashed curve shows the median
expectations in an ensemble of hypothetical experiments in which only
Standard Model background processes contribute, and the lower dotted curve
shows the median expectations in an ensemble of hypothetical
experiments in which a signal is also present.
The dark-shaded  band indicates the 68\% probability region centred on the
median signal+background expectation, while the light-shaded band indicates
the 95\% probability region.}
\end{figure}

\begin{figure}[p]
\centerline{\epsfig{file=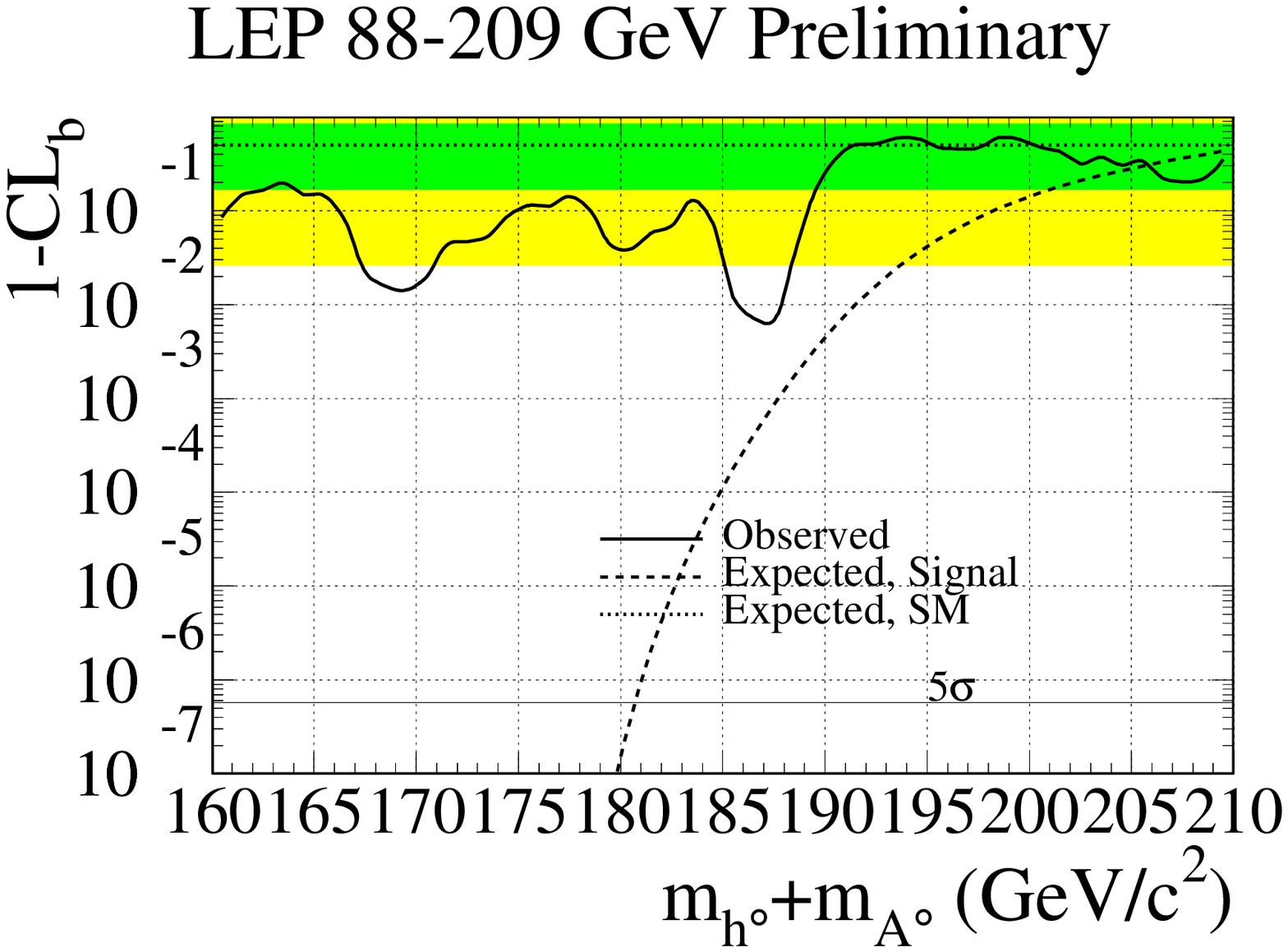,width=0.7\textwidth}}
\vspace{-5mm}
\centerline{\epsfig{file=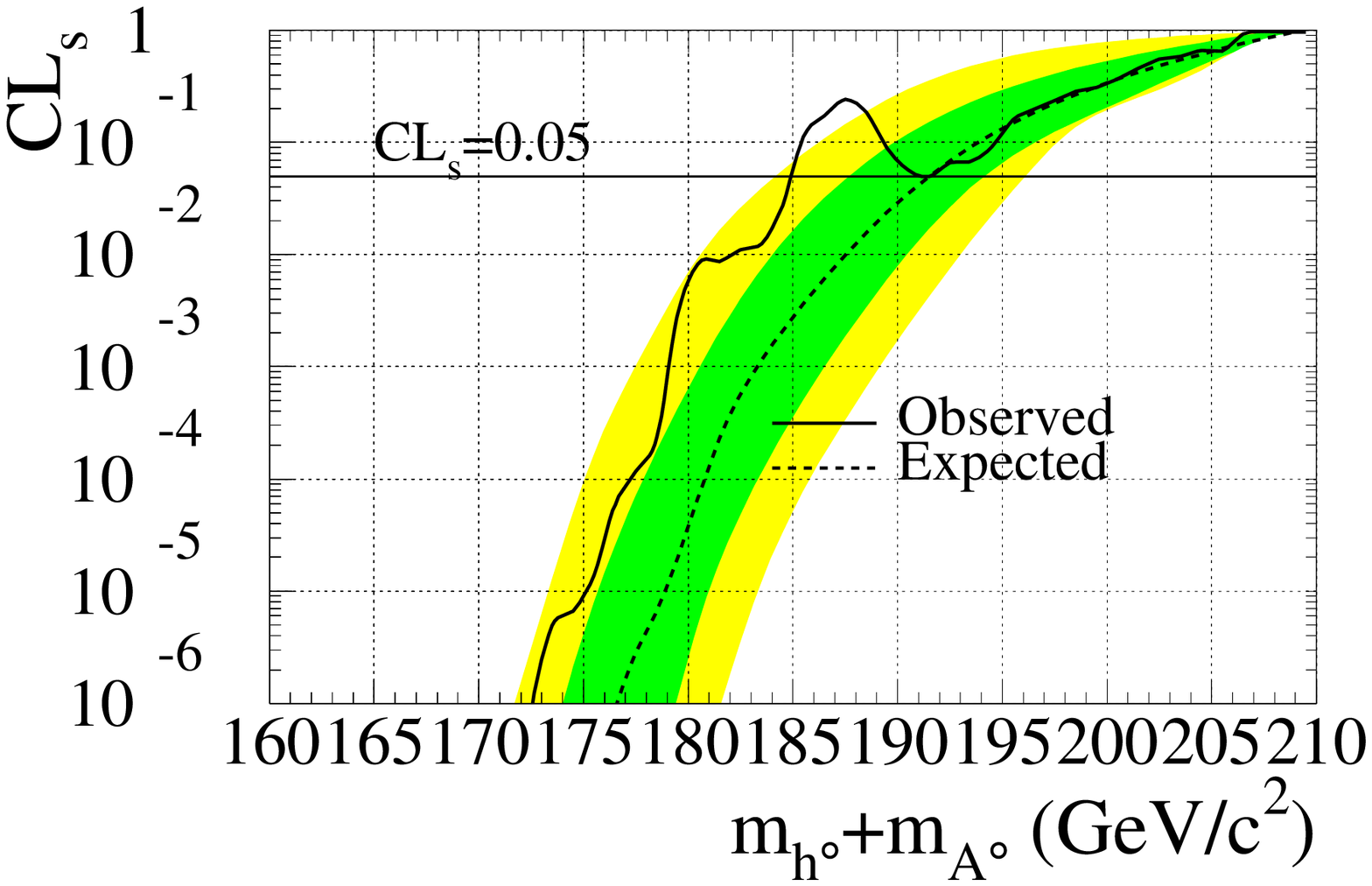,width=0.7\textwidth}}
\caption[]{\label{fig:diag}  
The values of $1-\clb$ and \cls\ for the 
$\mh\approx\mA$ diagonal 
($\tan\beta>20$, $\cos^2(\beta-\alpha)\approx 1$) in the \mhmax\ scenario.
The top plot shows the observed value of $1-\clb$ as a function of
$\mh+\mA$ in this scenario, as well as its median expected value (dashed line)
in the presence of a signal at the test mass.
The value of $1-\clb$ is expected to be uniformly
distributed between zero and one if there is no signal present.  The dark shaded
band is the 68\% probability region centred on $1-\clb=0.5$, and the
light-shaded band is the 95\% probability region centred also on $1-\clb=0.5$.
The solid line labelled ``5$\sigma$'' is drawn at $1-\clb=5.7\times 10^{-7}$.
In the lower plot, the observed value of \cls\ is shown in the same scenario
for the same high-$\tan\beta$ models.  The median expected \cls\ in an ensemble
of background-only experiments is shown with
a dashed line, and 68\% and 95\% probability contours are shown with dark and light
shading, respectively.  Models with $\cls < 0.05$ are excluded at the 95\% confidence
level.  The lowest unexcluded values of \mh\ and \mA\ correspond to models with
lower values of $\tan\beta$, for which $\mA\not=\mh$.}
\end{figure}

A more detailed view of the combined \ee\ra\ho\Ao\ search results is shown in
Figures~\ref{fig:diag2lnq} and ~\ref{fig:diag}.
In these figures, the values of $-2\ln Q$, $1-\clb$ and \cls\ are shown
for $\mh\approx\mA$ and at \tanb=20, as functions of \mh+\mA.  For these models,
$\cos^2(\beta - \alpha)\approx 1$, and the \ee\ra\ho\Zo\ searches do not contribute.
The significance of the excesses with $\mh\approx\mA$ in Figure~\ref{fig:mssm_clb}
are seen in the plot of $1-\clb$.  The quantity \cls\ is used to exclude the signal
hypothesis as a function
of the MSSM parameters.  The lowest unexcluded values of \mh\ and \mA\ correspond
to models with lower values of $\tan\beta$, for which $\mA\not=\mh$, and so these lower
bounds cannot be determined from Figure~\ref{fig:diag}.

\begin{figure}[p]
\centerline{
\epsfig{file=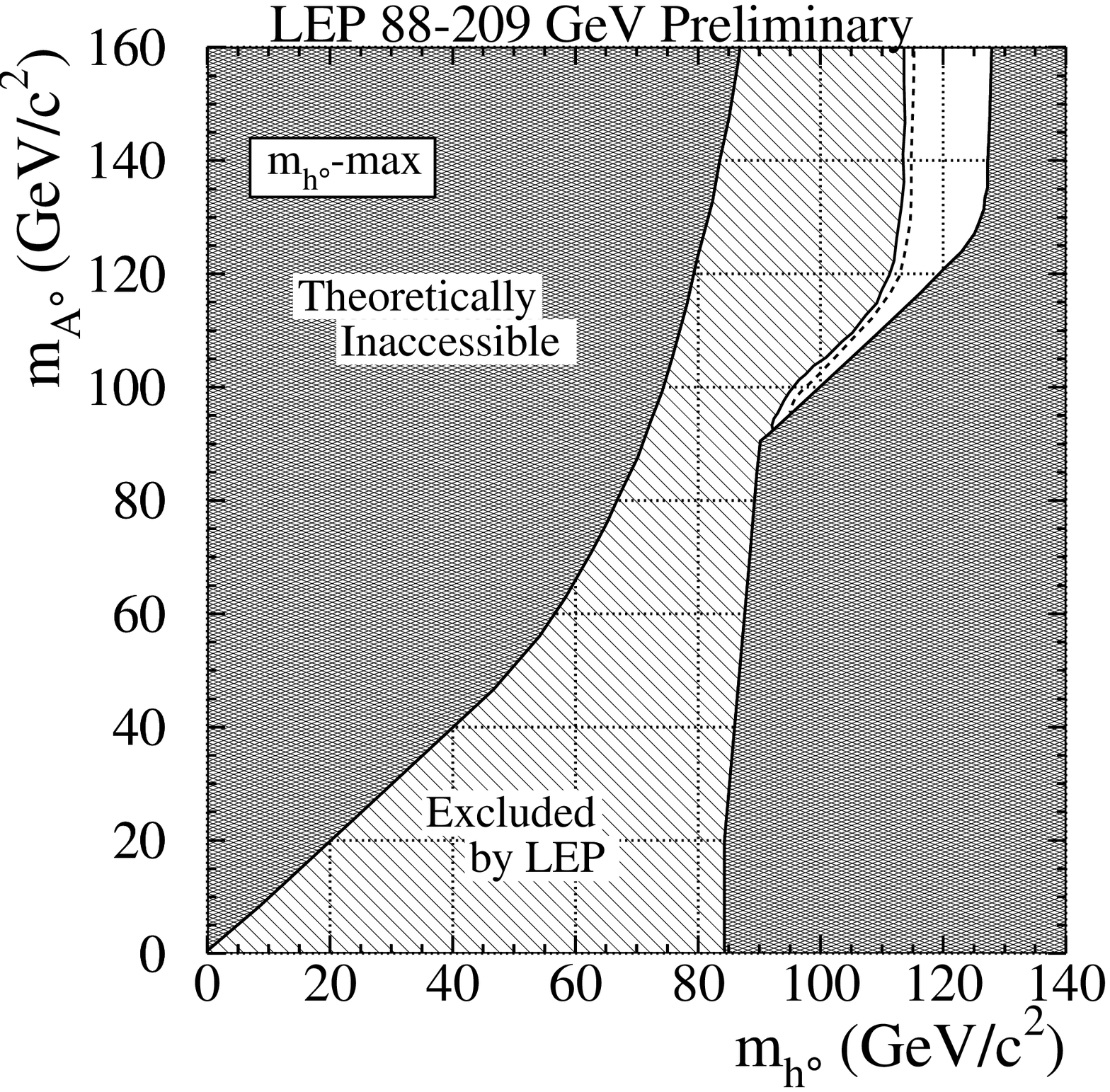,width=0.45\textwidth}
\epsfig{file=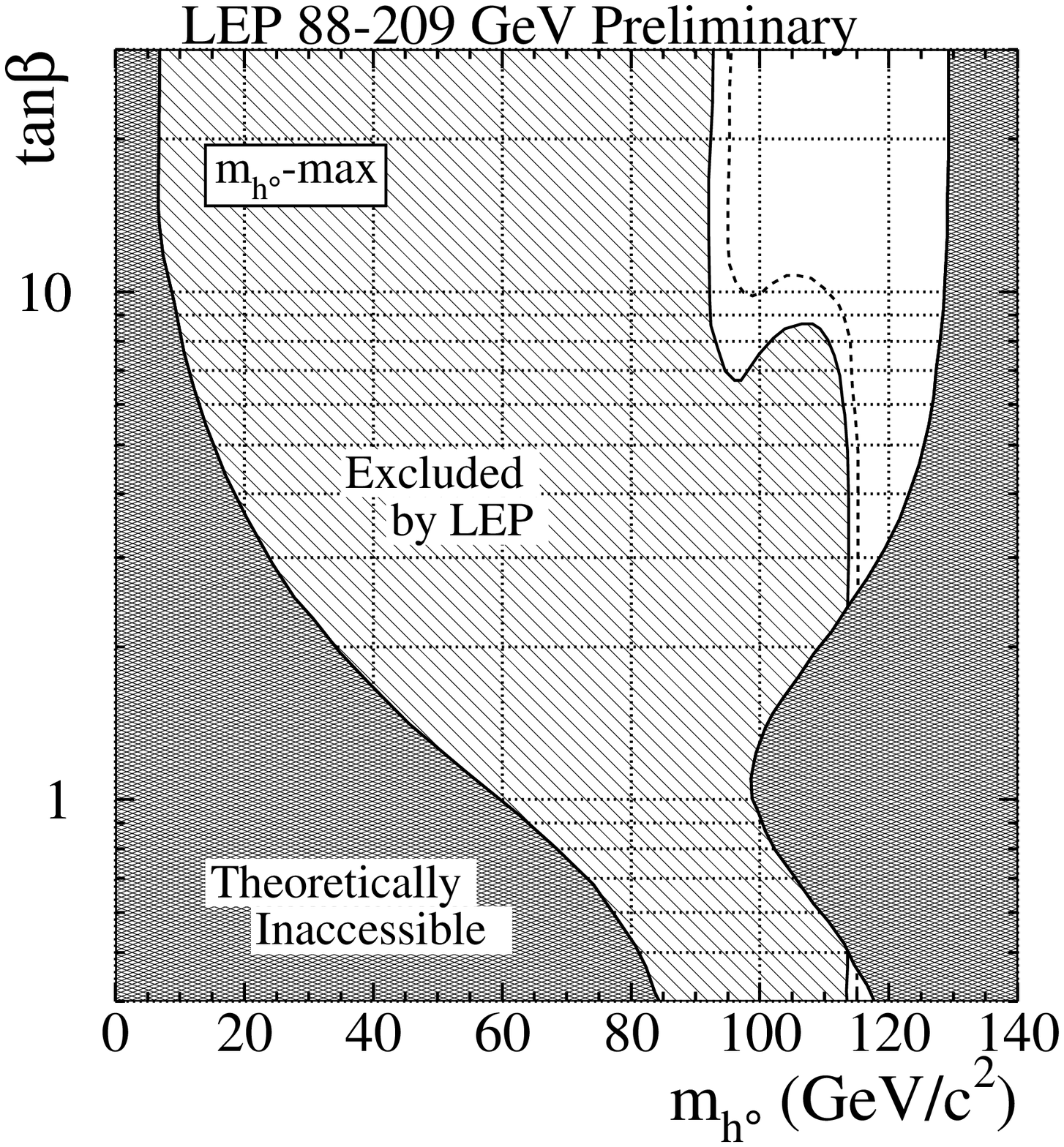,width=0.45\textwidth}
}
\centerline{
\epsfig{file=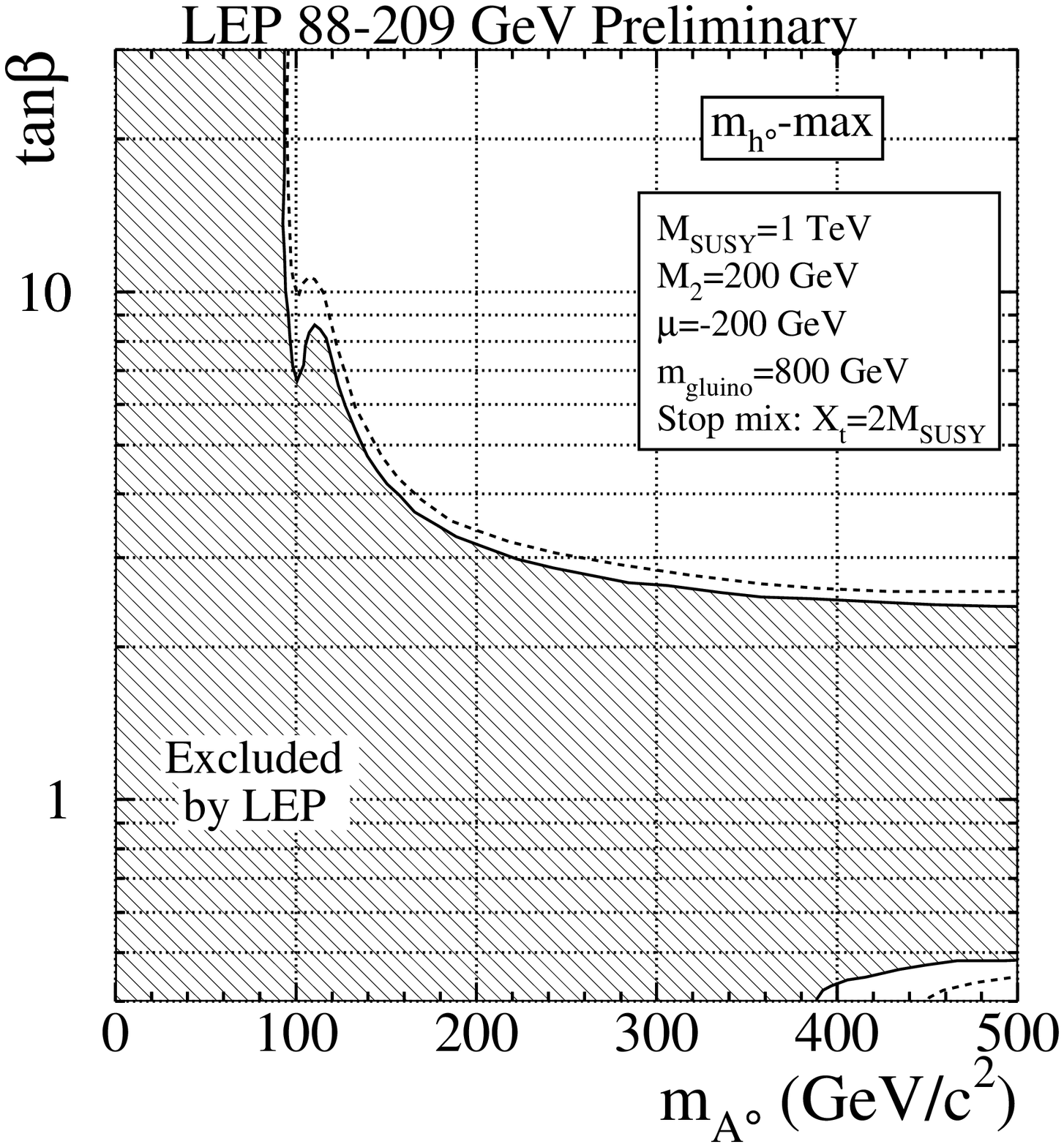,width=0.45\textwidth}
\epsfig{file=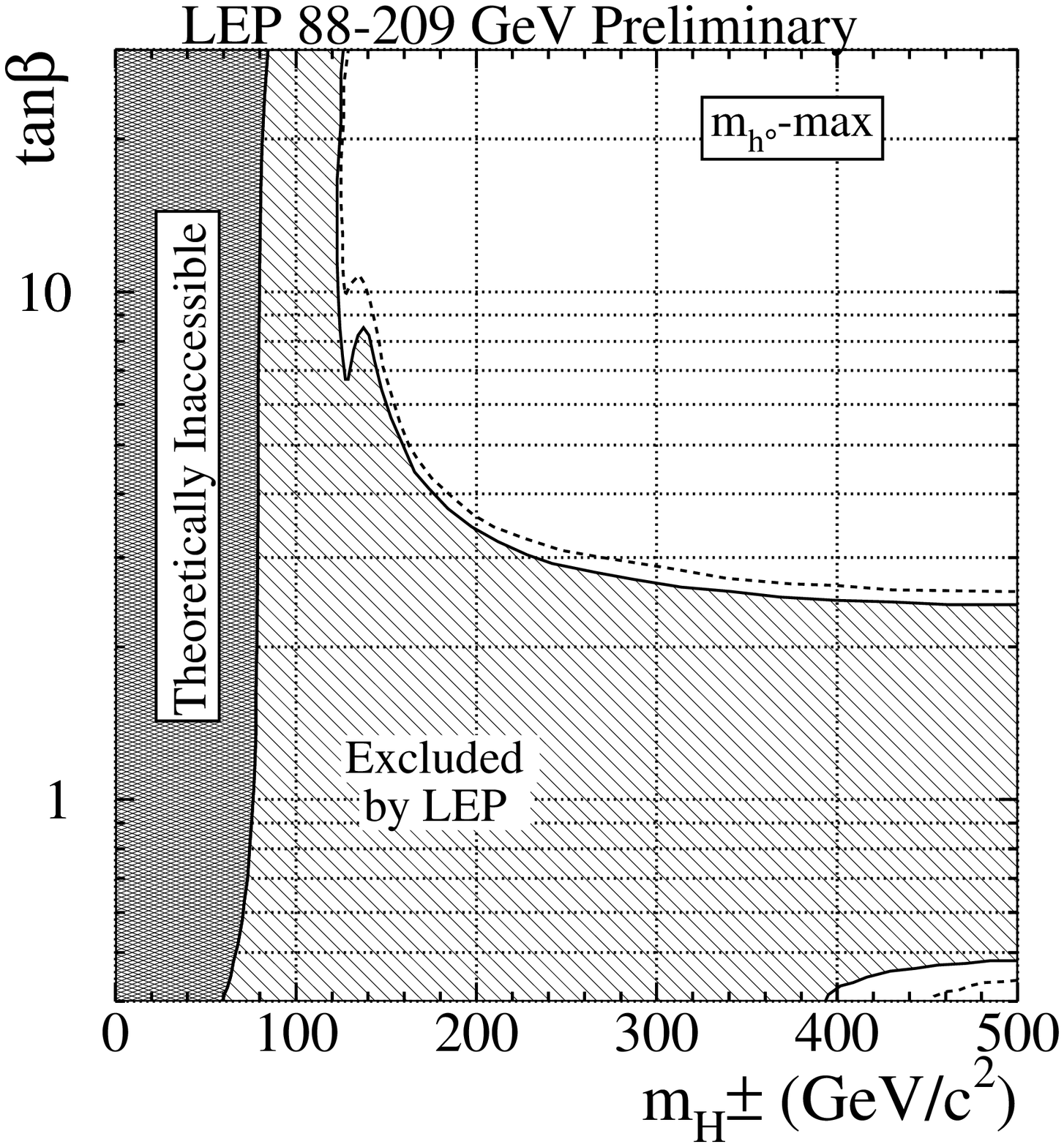,width=0.45\textwidth}
}
\caption[]{\label{fig:maxmh}
  The MSSM exclusion for the \mhmax\ benchmark scenario
  described in the text
  of Section~\ref{sect:mssmlimit}.
 This figure shows the excluded (diagonally hatched) 
  and theoretically disallowed (cross-hatched) regions as functions of the MSSM
         parameters in four projections:
         (upper left) the (\mh,~\mA) plane,
         (upper right) the (\mh,~\tanb) plane, 
         (lower left) the (\mA,~\tanb) plane and
         (lower right) the (\mHpm,~\tanb) plane.
         The dashed lines indicate the boundaries of the
         regions expected to be excluded at the 95\% CL if only SM background
         processes are present.
}
\end{figure}

\begin{figure}[p]
\centerline{
\epsfig{file=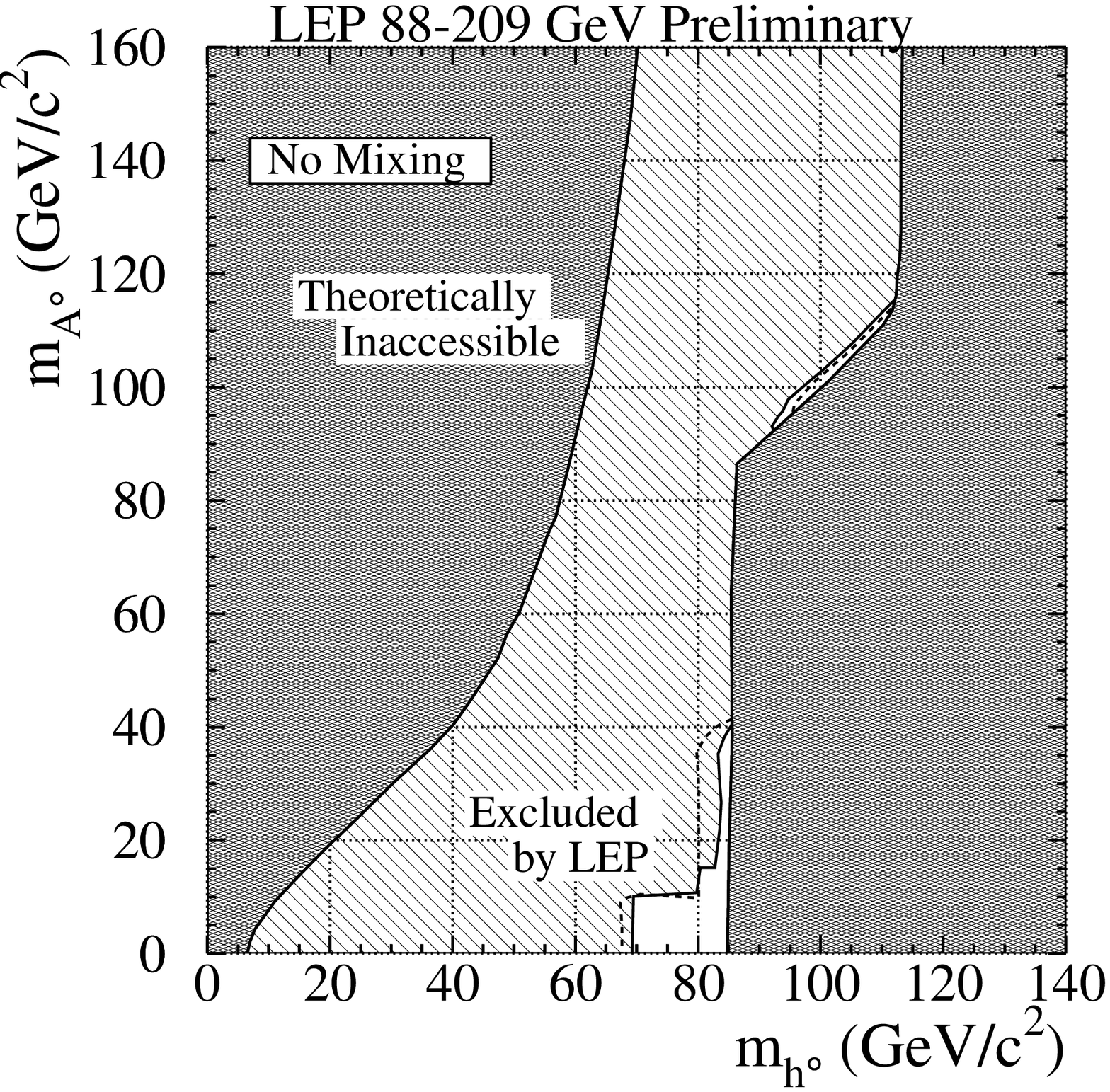,width=0.45\textwidth}
\epsfig{file=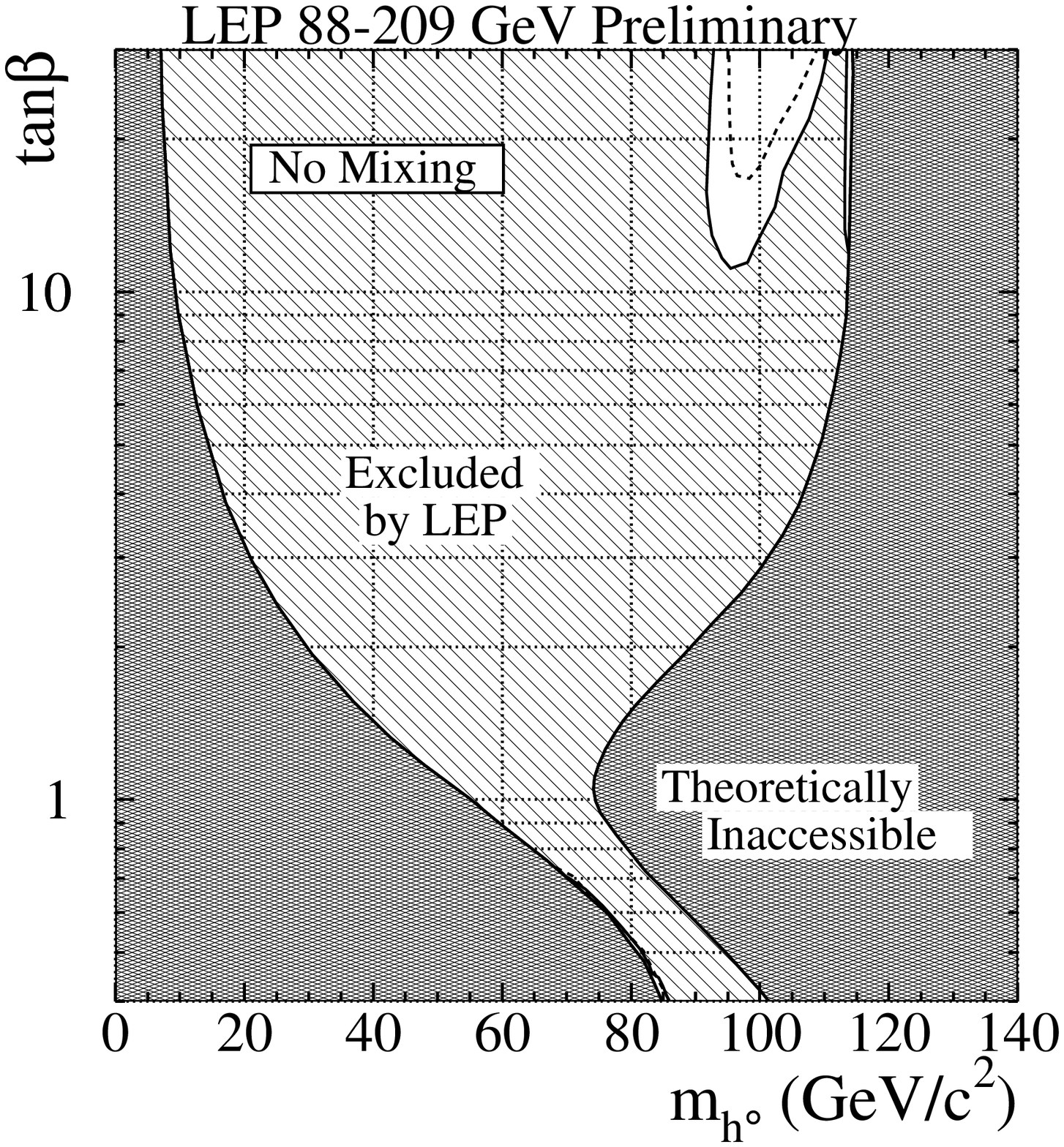,width=0.45\textwidth}
}
\centerline{
\epsfig{file=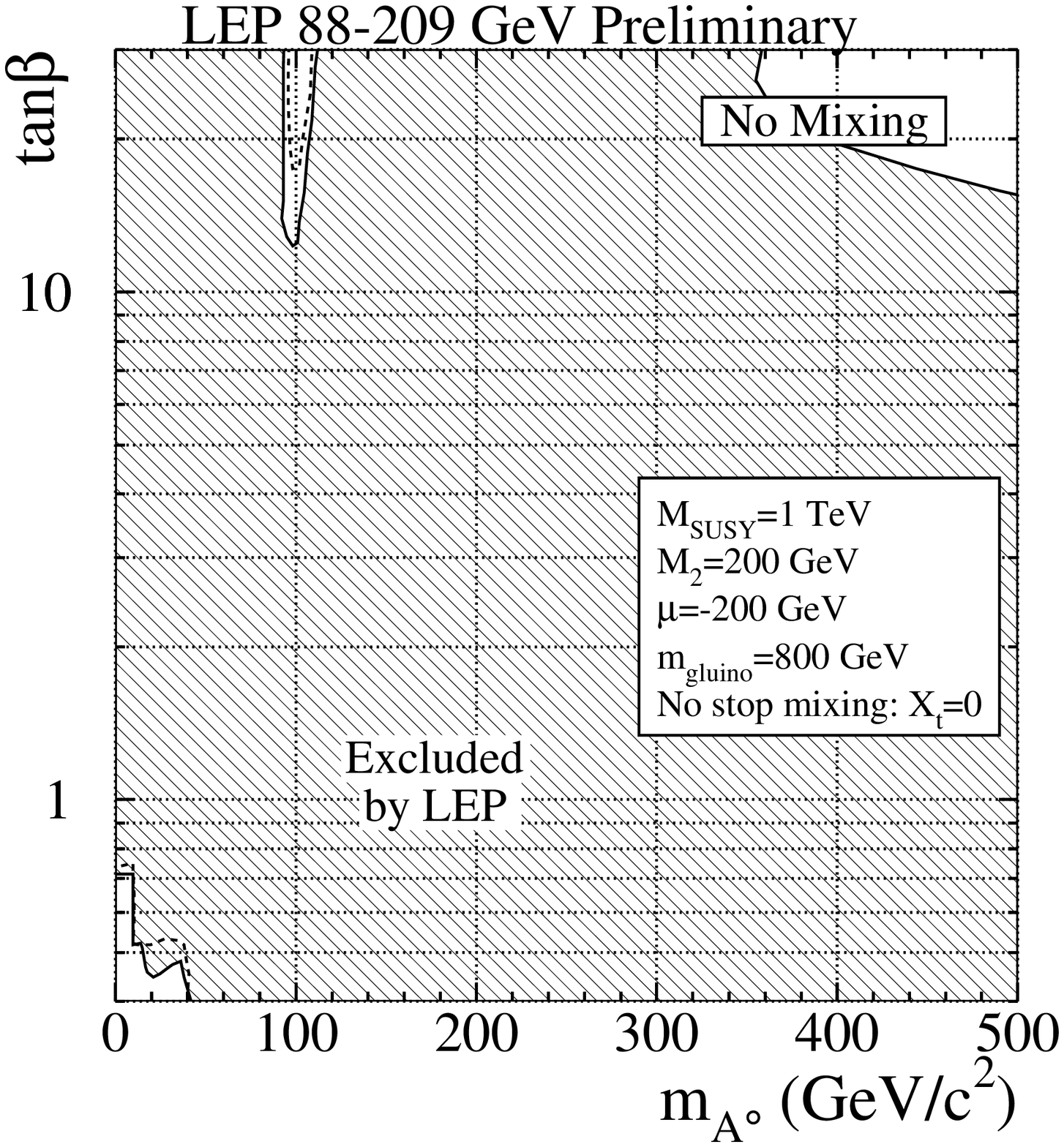,width=0.45\textwidth}
\epsfig{file=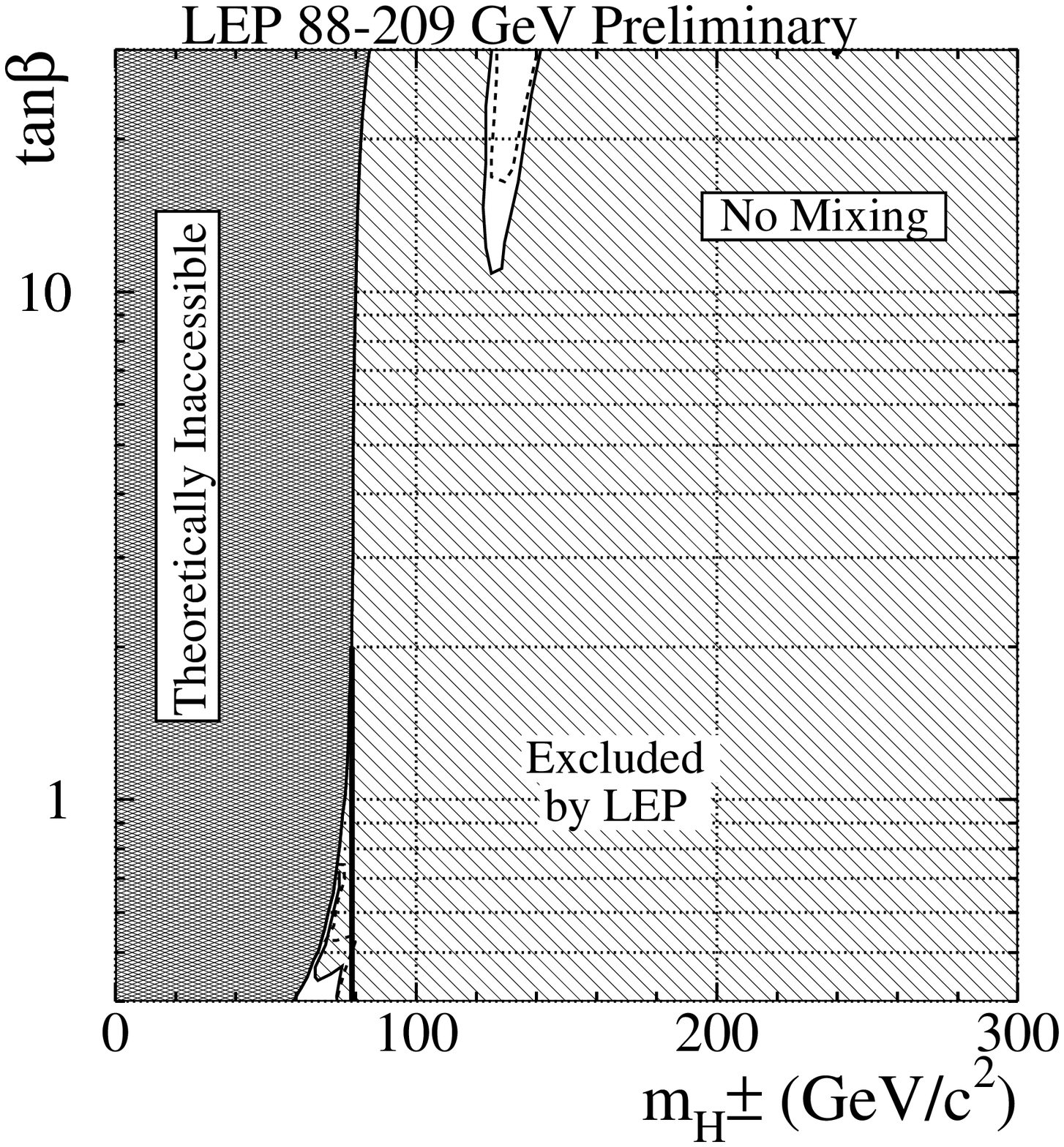,width=0.45\textwidth}
}
\caption[]{\label{fig:nomix}
  The MSSM exclusion for the ``no mixing'' benchmark scenario
  described in the text
  of Section~\ref{sect:mssmlimit}.  This figure shows the excluded (diagonally hatched)
  and theoretically inaccessible (cross-hatched) regions as functions of the MSSM
         parameters in four projections:
         (upper left) the (\mh,~\mA) plane,
         (upper right) the (\mh,~\tanb) plane,
         (lower left) the (\mA,~\tanb) plane and
         (lower right) the (\mHpm,~\tanb) plane.
         The dashed lines indicate the boundaries of the
         regions expected to be excluded at the 95\% CL if only SM background
         processes are present.   In the (\mHpm,~\tanb) projection,
         a dark vertical line is drawn at \mHpm=78.6~\gevcs, the lower bound
         obtained from direct searches at LEP.  Due to the decays \hwa, however,
         models with $\mHpm<74$ may not be excluded by the direct searches.  More
         study is needed to make a quantitative estimation of the impact of the
         \Hpm\ searches on this scenario.
}
\end{figure}

The 95\% CL exclusion contours are shown 
in Figure~\ref{fig:maxmh} for the \mhmax\ scenario, and
in Figure~\ref{fig:nomix} for the no-mixing scenario.  The results for the
large~$\mu$ scenario are discussed separately below.
In the no-mixing and \mhmax\ scenarios, limits are shown in
four projections: the (\mh,~\mA ) projection,
the (\mh,~\tanb ) projection, the (\mA,~\tanb ) projection,
and the (\mHpm,~\tanb ) projection.  

The observed and expected limits for \mh\ and \mA\ for the \mhmax\ and no-mixing
scenarios are given in Table~\ref{tab:mssmmhmalimits}.  
For the no-mixing scenario, the lower bounds on
\mh\ and \mA\ are given for $\tan\beta>0.7$ to highlight the search
sensitivity to heavy Higgs bosons.
For $\tan\beta < 0.7$, there is an unexcluded region
with \mA\ below 40~\gevcs\  and \mh\ above
65~\gevcs.  This region is
unexcluded  because the \ee\ra\ho\Zo\ra\Ao\Ao\Zo\ process dominates,
and Br(\Ao\ra\bb) is suppressed,
either kinematically, when $\mA < 10$~\gevcs, or because the coupling of the \Ao\ to \bb\ 
becomes sufficiently suppressed so that exclusion via b-tagging channels becomes
impossible.  For unexcluded models in the no-mixing scenario with $\tan\beta<0.7$, the 
mass of the charged Higgs boson is less than 74~\gevcs.  The lower bound obtained by the
combination of direct searches at LEP~\cite{LEPHIGGSCHARGED} is 78.6~\gevcs.
The LEP charged Higgs boson searches assume however that
Br(\Hp\ra\csbar )+Br(\Hp\ra\tpnu)=1.  This assumption is broken by
Br(\hpwa ), which can be as large as 0.6 for $\tan\beta=0.7$ and \mHpm=74~\gevcs, at the
extremum of the unexcluded area.  The decays of both the \Hp\ and the \Hm\ have
to be considered in signal events. 
The LEP-combined limits on the cross-section assuming only fermionic
\Hpm\ decays are of the order of 20\% of the predicted cross-section for \mHpm=74~\gevcs,
and so it is not clear that the entire unexcluded region can be covered by 
the constraint from charged Higgs boson searches.  Additional study is 
required to quantify the effect of the charged Higgs searches on this scenario.

For models with $\mA<4$~\gevcs, the decay branching fractions of the \Ao\ are uncertain.
In the \mhmax\ scenario for all values of \mh, and for the no-mixing scenario for
$\mh<65$~\gevcs, however, 
models with $\mA<4$~\gevcs\  are excluded regardless of the \Ao\ decay modes
because the production cross-section for \ho\Zo\ multiplied by Br(\ho\ra\bb ) provides
a sufficient signal to exclude these models.  If \mh\ is too low to allow
decays to \bb, then the additional width to the \Zo\ resonance from the \ho\Zo\
process or the \ho\Ao\ process would exceed the upper limit on the excess 
\Zo\ width~\cite{LEPEWWG2001}.


\begin{table}[htbp]
\begin{center}
\begin{footnotesize}
\begin{tabular}{|c|c|c|c|} \hline
Scenario & \mh\ limit (\gevcs ) & \mA\ limit (\gevcs ) & Excluded $\tan\beta$  \\
  & &  & observed limit (expected limit) \\\hline
\mhmax\   & 91.0 (94.6) & 91.9 (95.0) & $0.5<\tan\beta<2.4$ $(0.5<\tan\beta<2.6)$ \\
No Mixing & 91.5 (95.0) & 92.2 (95.3) & $0.7<\tan\beta<10.5$  $(0.8<\tan\beta<16.0)$ \\\hline
\end{tabular}
\end{footnotesize}
\vspace{0.5cm}
\caption{\label{tab:mssmmhmalimits} 
Limits on \mh\ and \mA\ in the \mhmax\ and no-mixing
benchmark scenarios explained in the text.  The 
median expected limits in an ensemble
of SM background-only experiments are listed in parentheses.  To highlight the
sensitivity of the searches for massive Higgs bosons, 
the limits on \mh\ and \mA\ are given
with the additional constraint of $\tan\beta>0.7$ for the no-mixing scenario.
If $\tan\beta$ is explored in the full region
to 0.4, then values of \mA\ below 40~\gevcs\  are not excluded 
for values of \mh\ above
65~\gevcs\  in the no-mixing scenario.  The excluded regions for the \mhmax\ and
no-mixing scenarios are shown in Figures~\protect\ref{fig:maxmh} and \protect\ref{fig:nomix}.}
\end{center}
\end{table}

The searches presented here allow regions of $\tan\beta$ to be excluded within
the contexts of the \mhmax\ and no-mixing scenarios.  For the \mhmax\ scenario,
values of \tanb\ between 0.5 and 2.4 are excluded, while
for the no mixing scenario,
values of \tanb\ between 0.7 and 10.5 are excluded.  The $\tan\beta$ limits in the 
\mhmax\ scenario are determined by the exclusion limit for the \ho\Zo\ process, which
depends strongly on the centre-of-mass energies LEP achieved.  For the no-mixing
scenario, the $\tan\beta$ limits are more complex.  The lower limit is determined
by the lack of sensitivity to the processes \ee\ra\ho\Zo\ra\Ao\Ao\Zo, where the
\Ao\ does not decay to \bb\ either because \tanb\ is too small or because the \Ao\ is too
light.  The upper limit is determined by the kinematic 
sensitivity of the \ee\ra\ho\Ao\ searches which leave an unexcluded 
region for $90 < \mA < 120$ GeV.  In this reigon, where Higgs-strahlung is suppressed by
the small value of $\sin^2(\beta-\alpha )$ and \ee\ra\ho\Ao\ is kinematically out of reach,
the maximum value of \mH\ is 114.6 GeV.  However, including the sensitivity to \ee\ra\Ho\Zo\ 
production does not improve the limits because models with \mH\ between 114.1~GeV and
114.6~GeV exist near the limits that are set.  The uncertainty on the model also
does not encourage the use of the Heavy Higgs signal to place limits on \tanb.

The region with $\mA > 300$ GeV in the no-mixing scenario also has an unexcluded portion
at high \tanb\ due to the fact that the limit in the Standard Model (\ho\Zo ) searches
is 114.1~GeV~\cite{LEPHIGGSEPS2001SM}, while the maximum possible value of \mh\ in this
scenario is 114.3~GeV.  This region at high \mA\ does not contribute to the \tanb\ limit
because it is at higher \tanb\ than the region at lower \mA.  If the model were to allow
even slightly higher maximal values of \mh\ in the no-mixing scenario, then the upper
value of the \tanb\ limit could be reduced because unexcluded regions will appear
at high \mA ; it is for this reason that the upper limit
on \tanb\ in this scenario is rather uncertain.

In a more general scan, where the MSSM parameters
are varied independently and the top quark mass is allowed to be
larger, the limits on \mh, \mA~and $\tan\beta$ are weaker
(see the discussions, for example, of Ref.~\cite{pr189}).  In particular, if the mass of the
top quark is 179~\gevcs\  (roughly 1$\sigma$ higher than measured central value),
then \tanb\ can no longer be excluded above 1.9 or below 0.6 in the \mhmax\ scenario.

\subsubsection{The {\boldmath Large~$\mu$\unboldmath} Scenario}
\label{sect:reslargemu}

The combination of the four experiments' results in the large~$\mu$ scenario
now includes the flavour-independent \ee\ra\ho\Zo\ searches with 
\ho\ra hadrons~\cite{ALEPHFLAVINDEP,DELPHIFLAVINDEP,L3FLAVINDEP,OPALMSSMMORIOND,OPALFLAVINDEP,LEPFLAVINDEP}.
In all cases, the cross-sections and decay branching ratios are computed with
SUBHPOLE2~\cite{MSSMMHBOUND5}.  Because of the large overlaps of the accepted signals, backgrounds,
and the observed candidates between the flavour-independent searches with the corresponding
b-tagged searches, only one set of searches is considered for each set of parameters
in the scenario, choosing between the flavour-independent set and the b-tagged set the channels
which yield the best expected sensitivity, given by the smallest median \cls\ in the
background-only hypothesis.
The previous combination~\cite{LEPHIGGSMSSMMORIOND} of the LEP search results, which
only included channels relying\footnote{The ALEPH leptonic and missing-energy
Standard Model Higgs channels have some sensitivity to non-\bb\ Higgs boson decays, but they are
not optimised for the flavour-blind interpretations, and more sensitivity is needed to exclude some
models in the large~$\mu$ scenario.}  on \ho\ decays to \bb\ and to \tautau,
did not exclude some choices of \mA\ and \tanb.  For the unexcluded models,
the signal events, while plentiful, were not selected because the leading decay branching ratios
of the \ho\ are to \cc, gluons and \WW; the \bb\ decays are suppressed by the choice
of model parameters and there is an insufficient branching ratio of the \ho\ to \tautau.

Because the light Higgs boson \ho\ has a mass less than 108~\gevcs\ for all choices
of (\mA,\tanb) in this model, and because \ee\ra\Ho\Zo\ is within kinematic reach
whenever \ee\ra\ho\Zo\ is suppressed by a small $\sin^2(\beta-\alpha)$ and 
$\mh+\mA>\sqrt{s}$, there is always a Higgs signal with sizeable strength
for all considered models within the large~$\mu$ scenario.  The challenge
of this scenario is to test models with non-\bb\ and non-\tautau\ decay modes.

A careful scan over the model space indicates that the addition of the flavour-indepen\-dent
\ee\ra\ho\Zo\ searches adds enough sensitivity to exclude the models which were
previously unexcluded, although for some model points they are interpreted as flavour-independent
\ee\ra\Ho\Zo\ searches.  This scenario is therefore entirely excluded at the 95\% confidence level.

\subsection{Coupling Strength Limits}

Searches for \ho\Ao\ production with reduced cross-sections or branching ratios
compared to those predicted
in the MSSM scenarios investigated here are of great interest.
More stringent limits on the \ho\Ao\ production cross-section allow tests of
models which either predict lower cross-sections or reduced branching ratios
of the Higgs bosons to the final states which are sought at LEP.  Examples
of such models are those involving substantial CP-violation in the MSSM Higgs
sector~\cite{cpviol} and general two-Higgs-doublet models (2HDMs) without 
SUSY constraints~\cite{higgshuntersguide}.

\begin{figure}[p]
\centerline{
\epsfig{file=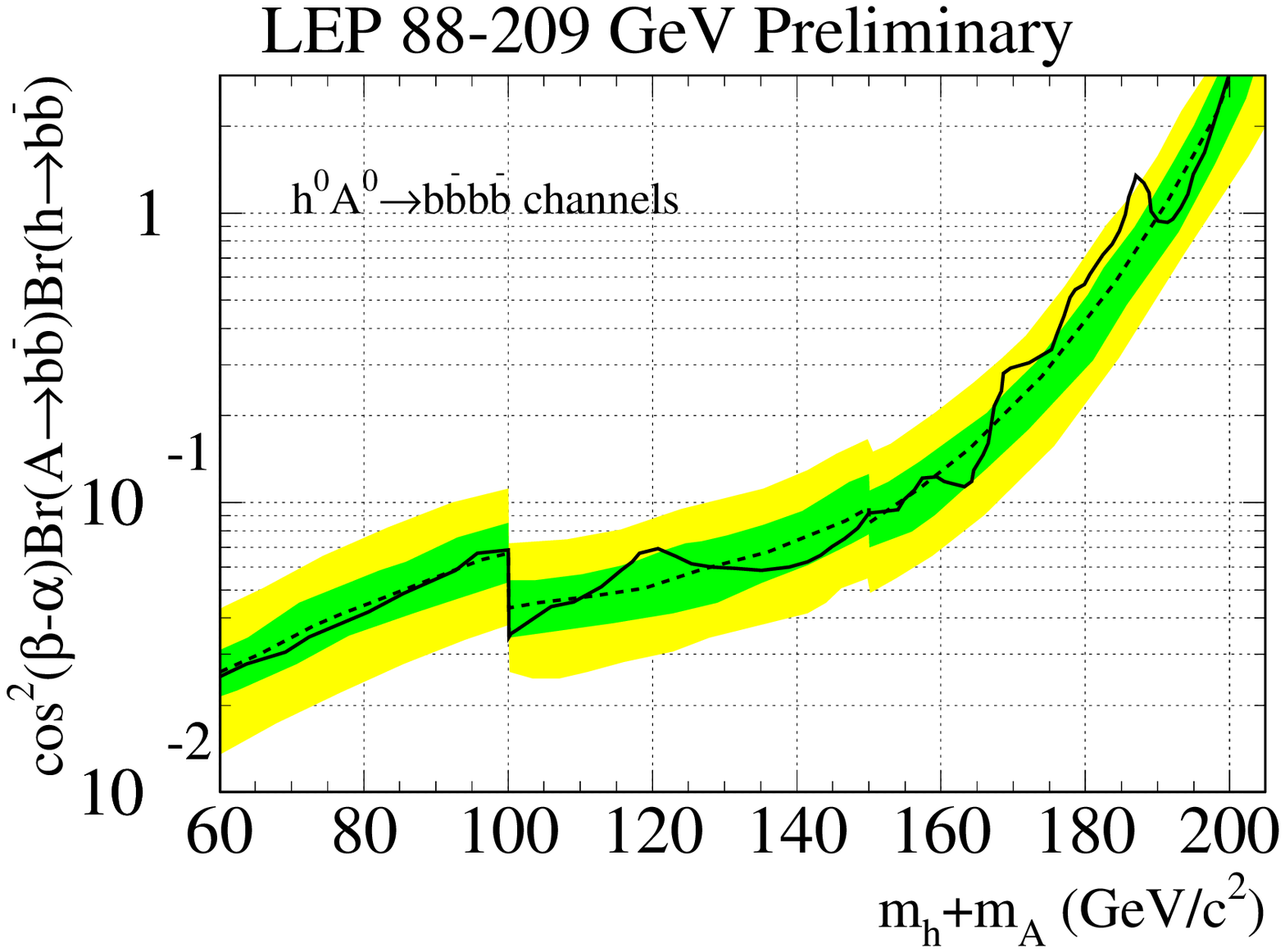,width=0.85\textwidth}
}
\caption[]{\label{fig:coslimitbbbb}
Limit on $\cos^2(\beta-\alpha){\mathrm {Br}}(\ho\ra\bb ){\mathrm {Br}}(\Ao\ra\bb )$, assuming $\mh\approx\mA$, 
and the energy-dependence of the \ee\ra\ho\Ao\ cross-section
from the \mhmax\ scenario.  The solid line is the observed
limit, and the dashed line is the median expected limit in an ensemble of 
hypothetical experiments in the absence of a signal.
Contours indicating the 68\% and 95\% probability bands centred on the median expectation
show the expected variation of the limit in an ensemble of background-only
experiments.
}
\end{figure}

\begin{figure}[p]
\centerline{
\epsfig{file=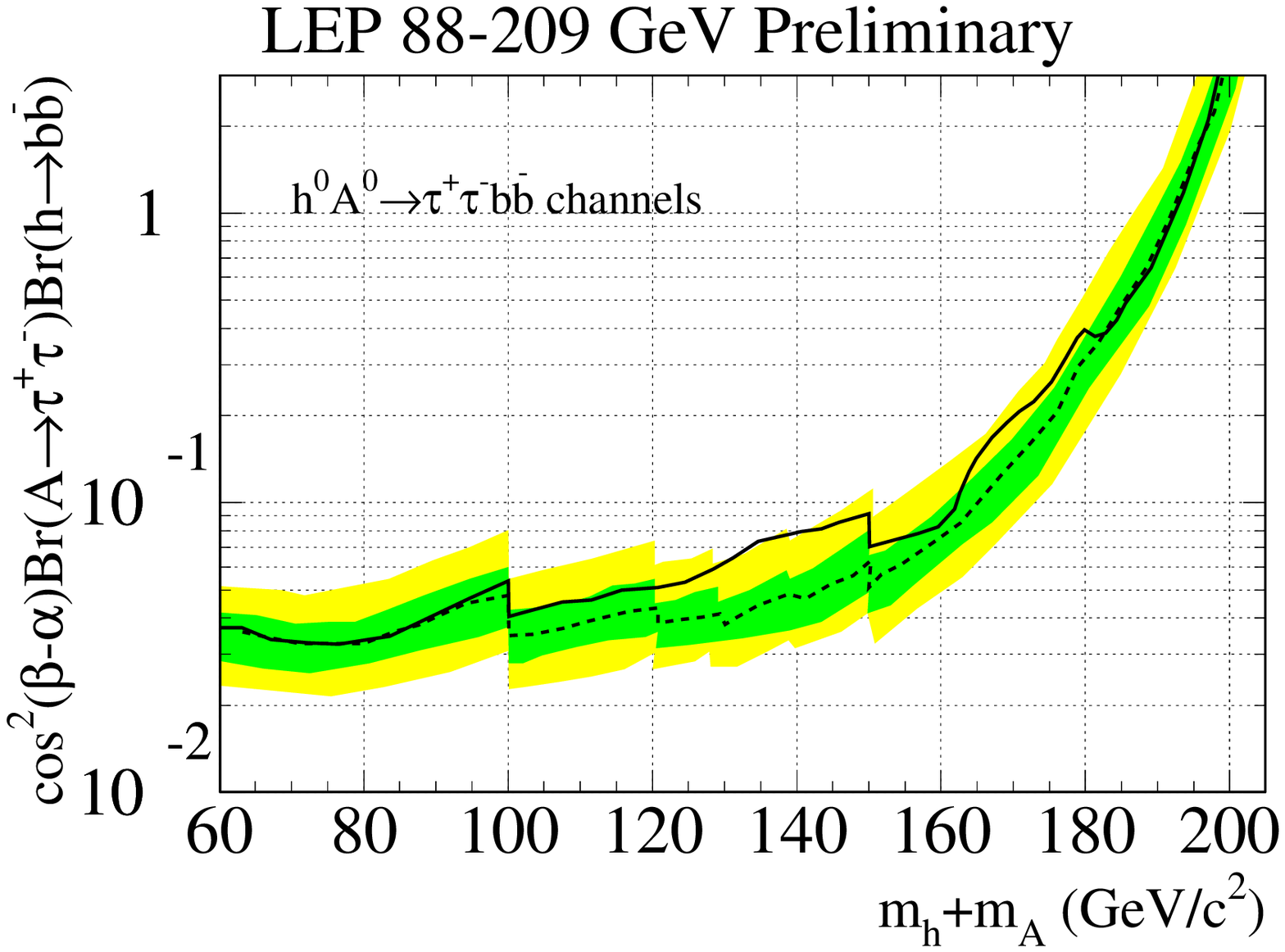,width=0.85\textwidth}
}
\caption[]{\label{fig:coslimitbbtatau}
Limit on $\cos^2(\beta-\alpha){\mathrm {Br}}(\ho\ra\bb ){\mathrm {Br}}(\Ao\ra\tautau )$,
assuming $\mh\approx\mA$
and the energy-dependence of the \ee\ra\ho\Ao\ cross-section
from the \mhmax\ scenario.  The solid line is the observed
limit, and the dashed line is the median expected limit in an ensemble of 
hypothetical experiments in the absence of a signal.
Contours
indicating the 68\% and 95\% probability bands centred on the median expectation
show the expected variation of the limit in an ensemble of background-only
experiments.
}
\end{figure}

\begin{figure}[p]
\centerline{
\epsfig{file=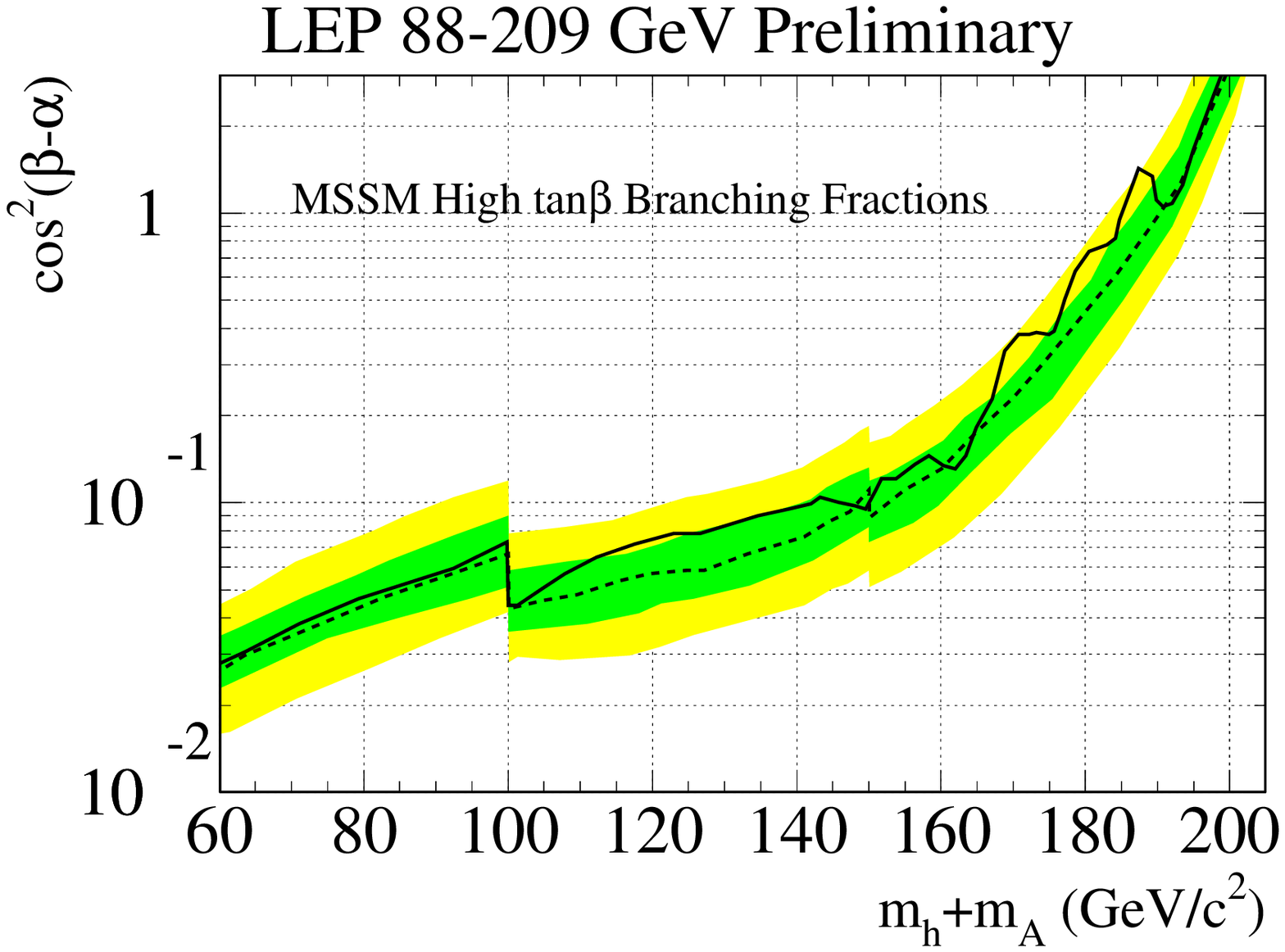,width=0.85\textwidth}
}
\caption[]{\label{fig:coslimitmhmax}
Limit on $\cos^2{\beta-\alpha}$, assuming $\mh\approx\mA$,
and the fixed branching fractions 
Br(\ho\ra\bb )=0.94, Br(\Ao\ra\bb )=0.92, 
Br(\ho\ra\tautau )=0.06 and Br(\Ao\ra\tautau )=0.08, typical of the \mhmax\ scenario
for values of \tanb\ greater than 10.  The solid line is the observed
limit, and the dashed line is the median expected limit in an ensemble of 
hypothetical experiments in the absence of a signal.  
Contours
indicating the 68\% and 95\% probability bands centred on the median expectation
show the expected variation of the limit in an ensemble of background-only
experiments.
}
\end{figure}

The \ee\ra\ho\Ao\ search results from the four experiments are combined using 
the MSSM model considered in the \mhmax\ scenario above to compute the dependence
of the \ee\ra\ho\Ao\ cross-section on the centre-of-mass energy. 
The coupling limits are produced separately for the set of \ho\Ao\ra\bb\bb\ search
channels, 
for the set of \ho\Ao\ra\bb\tautau\ search channels, and also for a combination of all
\ho\Ao\ search channels assuming fixed branching ratios.  The branching ratios
chosen for the third set of coupling limits is
Br(\ho\ra\bb )=0.94, Br(\Ao\ra\bb )=0.92, 
Br(\ho\ra\tautau )=0.06 and Br(\Ao\ra\tautau )=0.08,
which are typical in the \mhmax\ scenario for values of \tanb\ greater than 10. 
In all cases, $\cos^2(\beta-\alpha)=1$, and the signal is multiplied by a scale
factor such that the scaled signal is excluded at exactly the 95\% confidence level
($\cls = 0.05$).  Presently, no \ho\Ao\ra\tautau\tautau\ searches are combined.
The coupling strength limits for the three combinations of channels
are shown in Figures~\ref{fig:coslimitbbbb} through~\ref{fig:coslimitmhmax}.
These limits can be interpreted as upper bounds on
$\cos^2(\beta-\alpha ){\mathrm Br}(\ho\ra\bb ){\mathrm Br}(\Ao\ra\bb )$,
$\cos^2(\beta-\alpha ){\mathrm Br}(\ho\ra\bb ){\mathrm Br}(\Ao\ra\tautau )$,
and $\cos^2(\beta-\alpha)$ assuming the fixed branching ratios mentioned above.
The limits for the \bb\tautau\ channels can be interpreted for either the
\ho\ or the \Ao\ decaying into \tautau, while the other decays to \bb.
The experimental limits are shown in the figures along with the limits
that are expected in an ensemble of hypothetical experiments in which there is
no signal.

\section*{Acknowledgements}
We congratulate our colleagues for the LEP Accelerator Division for the successful
running in the year 2000 at the highest energies, and would like to express our
thanks to the engineers and technicians in all our institutions for their 
contributions to the excellent performance of the four LEP experiments.
The LEP Higgs working group acknowledges the fruitful cooperation between the
experiments in exchanging the experimental results and developing and applying
procedures for combining them.



%
\clearpage
\newpage


\clearpage
\newpage

\begin{thebibliography}{99}
%
\bibitem{MSSMMHBOUND1} Y.~Okada, M.~Yamaguchi and T.~Yanagida, Prog. Theor. Phys. {\bf 85} (1991) 1.

\bibitem{MSSMMHBOUND2} J.~Ellis, G.~Ridolfi and F.~Zwirner, Phys. Lett. {\bf B257} (1991) 83.

\bibitem{MSSMMHBOUND3} H.E.~Haber and R.~Hempfling. Phys. Rev. Lett. {\bf 66} (1991) 1815.

\bibitem{MSSMMHBOUND4} M.~Carena, J.~R.~Espinosa, M.~Quir\'os and C.E.M.~Wagner, Phys. Lett. {\bf B355} (1995) 209.

\bibitem{MSSMMHBOUND5} M.~Carena, M.~Quir\'os and C.E.M.~Wagner, Nucl. Phys. {\bf B461} (1996) 407.

\bibitem{MSSMMHBOUND6} H.E.~Haber, R.~Hempfling and A.H.~Hoang, Zeit f\"ur Phys. {\bf C75} (1997) 539.

\bibitem{MSSMMHBOUND7} S.~Heinemeyer, W.~Hollik and G.~Weiglein, Eur. Phys. Jour. {\bf C9} (1999) 343.

\bibitem{MSSMMHBOUND8} J.~R.~Espinosa and R.~Zhang, Nucl. Phys. {\bf B586} (2000) 3.

\bibitem{ALEPHSMLETTER} ALEPH Collab., R.~Barate~\etal, 
Phys. Lett. {\bf B495} (2000) 1.

\bibitem{DELPHISMLETTER} DELPHI Collab., P.~Abreu~\etal, 
Phys. Lett. {\bf B499} (2001) 23.


\bibitem{L3SMLETTER} L3. Collab., ``Search for the Standard Model Higgs Boson with the L3
Experiment at LEP'', L3 Note 2688, June 2001.  Submitted to Phys. Lett.~B.

\bibitem{OPALSMLETTER} \OPALColl, 
published in Phys. Lett. {\bf B499} (2001) 38.

\bibitem{ALEPHMSSMMORIOND} ALEPH Collab., ``Searches for neutral Higgs bosons of the
MSSM at centre-of-mass energies up to 209~GeV with teh ALEPH detector at LEP'',
ALEPH 2001-022 CONF 2001-019, 6 March, 2001.


\bibitem{DELPHIMSSMEPS} DELPHI Collab., ``Searches for neutral supersymmetric Higgs bosons
in \ee\ collisions up to $\sqrt{s}=209$~GeV'', DELPHI 2001-070 CONF 498, 4 July, 2001.

\bibitem{L3MSSMMORIOND} L3 Collab., ``Search for Neutral Higgs Bosons of the Minimal Supersymmetric
Standard Model in \ee\ Interactions at $\sqrt{s}$ up to 209~GeV'', L3 Note 2692, July 2001.

\bibitem{OPALMSSMMORIOND} \OPALColl, ``Searches for Higgs Bosons in Extensions to the Standard Model
in \ee\ Collisoins at the Highest LEP Energies''
OPAL Physics Note PN472, 27 February, 2001.

\bibitem{ALEPHFLAVINDEP} ALEPH Collab., ``A flavour-independent search for the Higgsstrahlung
process in \ee\ collisions at centre-of-mass energies from 189 to 209~GeV'', ALEPH 2001-021 CONF 2001-018.

\bibitem{DELPHIFLAVINDEP} DELPHI Collab., ``Generalised Search for Hadronic Decays of
Higgs Bosons with the DELPHI Detector at LEP-2'', DELPHI 2001-070 CONF 498, July 2001.

\bibitem{L3FLAVINDEP} L3 Collab., ``Flavour Independent Search for Hadronically Decaying Higgs Boson
in Higgs-strahlung Process at $\sqrt{s}$ up to 209~GeV'', L3 Note 2693, July 2001.

\bibitem{OPALFLAVINDEP} OPAL Collab., Eur. Phys. J. {\bf C18} (2001) 425; \\
OPAL Collab., ``Model Independent Searches for Scalar Bosons with the OPAL Detector at LEP'',
Physics Note PN449, available at {\tt http://opal.web.cern.ch/Opal/pubs/physnote/info/pn449.html}.

\bibitem{LEPFLAVINDEP} The ALEPH, DELPHI, L3 and OPAL Collaborations,
and the LEP Higgs Working Group, ``Generalised Search for Hadronic Decays
of Higgs Bosons at LEP-2'', LEP Higgs WG Note 2001-07, July, 2001.

\bibitem{HZHA} HZHA generator: P.~Janot, in ``Physics at LEP2'',
               edited by G.~Altarelli, T.~Sj\"{o}strand and
               F.~Zwirner, CERN 96-01 Vol.~2  p.309. \\
For HZHA3 and HZHA2, see {\tt http://alephwww.cern.ch/$\sim$janot/Generators.html}.

\bibitem{weigheiholl}  
S.~Heinemeyer, W.~Hollik and G. Weiglein, Phys. Rev. {\bf D58} (1998) 091701, 
Phys. Lett. {\bf B440} (1998) 296, hep-ph/9807423 and 
JHEP~0006 (2000) 009.

\bibitem{feynhiggs} S.~Heinemeyer, W.~Hollik and~G. Weiglein, 
       Comp. Phys. Comm. {\bf 124} (2000) 76; Also see {\tt http://www.feynhiggs.de}.


\bibitem{LEPHIGGSEPS2001SM}  The ALEPH, DELPHI, L3 and OPAL Collaborations, and
the LEP Higgs Working Group, ``Search for the Standard Model Higgs Boson at LEP'',
LEP Higgs WG Note 2001-03, July 2001.

\bibitem{ref:cousinshighland} R.~D.~Cousins and V.~L.~Highland, Nucl.~Instr.~Meth
{\bf A320} (1992) 331.

\bibitem{LEPHIGGSCHARGED} ALEPH, DELPHI, L3 and OPAL Collaborations, and the LEP Higgs
Working Group, ``Search for Charged Higgs boson: Preliminary combined results using
LEP data collected at energies up to 209 GeV'' LEP Higgs WG Note 2001-05.

\bibitem{LEPHIGGS202} ALEPH, DELPHI, L3, OPAL Collab., and
the LEP working group for the Higgs boson searches, 
``Search for Higgs bosons: Preliminary combined results using LEP data 
collected at energies up to 202 GeV,'' CERN-EP/2000-055, 25 Apr 2000.

\bibitem{RPP2000} D.E. Groom {\it et al}, Eur. Phys. Jour. {\bf C15} (2000) 1,
available on the PDG WWW pages (URL: {\tt http://pdg.lbl.gov/}).
 
\bibitem{newbenchmarks} M.~Carena, S.~Heinemeyer, C.~E.~M.~Wagner and G.~Weiglein,
hep-ph/9912223.

\bibitem{carenamrennawagner} M.~Carena, S.~Mrenna and C.~Wagner, Phys. Rev. D60 (1999) 075010.

\bibitem{reconciliation} M.~Carena, H.~E.~Haber, S.~Heinemeyer, W.~Hollik, C.~E.~M.~Wagner
and G.~Weiglein, 
Nucl. Phys. {\bf B580} (2000) 29.

\bibitem{espinosareconciliation} J.~R.~Espinosa and R.-J.~Zhang, JHEP 0003:026 (2000).


\bibitem{pr189} 
\OPALColl, G.~Abbiendi \etal, 
Eur. Phys. J. {\bf C12} (2000) 567-586.

\bibitem{LHWGTAMPERE} The ALEPH, DELPHI, L3 and OPAL Collaborations, and the LEP Higgs 
Working Group, ``Searches for HIggs Bosons: Preliminary Combined Results from the 
Four LEP Experiments at $\sqrt{s}\approx 189$~GeV, 
ALEPH 99-081 CONF 99-052, DELPHI 99-142 CONF 327, L3 Note 2442, OPAL 
Technical Note TN614, July 1999.

\bibitem{LEPEWWG2001} The ALEPH, DELPHI, L3 and OPAL Collaborations, the LEP
Electroweak Working Group, and the SLD Heavy Flavour and Electroweak Groups,
``A Combination of Preliminary Electroweak Measurements and Constraints on
the Standard Model'', LEPEWWG/2001-01, May 2001.

\bibitem{LEPHIGGSMSSMMORIOND} ALEPH, DELPHI, L3 and OPAL Collaborations, and the LEP Higgs
Working Group, ``Searches for the Neutral Higgs Bosons of the MSSM: Preliminary
Combined Results using LEP Data Collected at Energies up to 209 GeV'', LHWG note 2001-02.

\bibitem{akeroyd} A.~G.~Akeroyd and W.~J.~Stirling, Nucl. Phys. {\bf B447} (1995), 3.


\bibitem{cpviol} M.~Carena, J.~Ellis, A.~Pilaftsis and C.~Wagner, Nucl. Phys. {\bf B586} (2000) 92.

\bibitem{higgshuntersguide} J.~Gunion, H.~Haber, G.~Kane and S.~Dawson, 
{\it The Higgs Hunter's Guide}, Addison-Wesley Publishing Company (1990).

\end{thebibliography}
\end{document}